\documentclass[aps,prb,twocolumn,amsmat,amssymb,amsfonts,superscriptaddress]{revtex4-1}
\usepackage{amsmath}
\usepackage{tikz}
\usetikzlibrary{arrows}
\usepackage{ifthen}

\usetikzlibrary{decorations.pathreplacing,decorations.markings}
\usepackage{times}
\usepackage{hyperref}

\usepackage{color}
\usepackage{graphics}
\usepackage{graphicx}
\usepackage[percent]{overpic}
\usepackage{wasysym}
\graphicspath{{../PRX/figures/}}

\newcommand{\tcRStoFK}{$\phi/\pi=1$}
\newcommand{\tcFKtoAG}{$\phi/\pi=0.883$}
\newcommand{\tcAGtoSNP}{$\phi/\pi=0.440$}
\newcommand{\tcSNPtoQ}{$\phi/\pi=0.428$}
\newcommand{\tcQtoQP}{$\phi/\pi=0.396$}
\newcommand{\tcQPtoFM}{$\phi/\pi=0.385$}
\newcommand{\tcFMtoAK}{$\phi/\pi=0.086$}
\newcommand{\tcAKtoSN}{$\phi/\pi=-0.155$}
\newcommand{\tcSNtoRS}{$\phi/\pi=-0.265$}
\newcommand{\AG}{A$\Gamma$}
\newcommand{\AK}{AK}
\newcommand{\FK}{FK}
\newcommand{\SN}{SN}

\tikzset{
  on each segment/.style={
    decorate,
    decoration={
      show path construction,
      moveto code={},
      lineto code={
        \path [#1]
        (\tikzinputsegmentfirst) -- (\tikzinputsegmentlast);
      },
      curveto code={
        \path [#1] (\tikzinputsegmentfirst)
        .. controls
        (\tikzinputsegmentsupporta) and (\tikzinputsegmentsupportb)
        ..
        (\tikzinputsegmentlast);
      },
      closepath code={
        \path [#1]
        (\tikzinputsegmentfirst) -- (\tikzinputsegmentlast);
      },
    },
  },
  mid arrow/.style={postaction={decorate,decoration={
        markings,
        mark=at position .5 with {\arrow[#1]{stealth}}
      }}},
}

\def\circledarrow#1#2#3{ 
\draw[#1,->] (#2) +(80:#3) arc(80:-260:#3);
}
\def\rs{45}
\def\chrlAK#1{%
  \pgfmathparse{int(#1)}%
  \ifnum\pgfmathresult>0
  \foreach \x in {0,...,#1}
     \filldraw (1.3*\x,-.1) circle (0.1);
  \foreach \x in {0,...,#1}
     \filldraw (1.3*\x,1.1) circle (0.1);
  \draw[line width=0.2mm] (-0.2,1.1) -- (8.0,1.1);
  \draw[line width=0.2mm] (-0.2,-.1) -- (8.0,-.1);
   \pgfmathtruncatemacro{\y}{#1}
  \foreach \x in {0,...,\y}
        \draw[line width=0.2mm] (1.3*\x,1.1) -- (1.3*\x,-.1);
  \foreach \x in {0,...,5} {
     \ifthenelse{\x=0 \OR \x=2 \OR \x=4}{
     \path [draw=red,line width=\rs*0.0023882976987468mm,postaction={on each segment={mid arrow=red}}]
        (1.3*\x+0.1,0) -- (1.3*\x+1.1,0) -- (1.3*\x+0.1,1)-- (1.3*\x+0.1,0);
      \draw (1.3*\x+0.4,0.35) node {-};
      \path [draw=blue,line width=\rs*0.0023882976987468mm,postaction={on each segment={mid arrow=blue}}]
         (1.3*\x+0.2,1.0) -- (1.3*\x+1.2,1.00) -- (1.3*\x+1.2,0.00) -- (1.3*\x+0.2,1.0);
      \draw (1.3*\x+0.85,0.65) node {+};
    } {
     \path [draw=purple,line width=\rs*0.0058263516163867mm,postaction={on each segment={mid arrow=purple}}]
        (1.3*\x+0.1,0) -- (1.3*\x+1.1,0) -- (1.3*\x+0.1,1)-- (1.3*\x+0.1,0);
      \draw (1.3*\x+0.4,0.35) node {-};
      \path [draw=blue,line width=\rs*0.0058263516163867mm,postaction={on each segment={mid arrow=blue}}]
         (1.3*\x+0.2,1.0) -- (1.3*\x+1.2,1.00) -- (1.3*\x+1.2,0.00) -- (1.3*\x+0.2,1.0);
      \draw (1.3*\x+0.85,0.65) node {+};
      }
    }
    \fi
  }

\def\ra{15}
\def\chrlalpha#1{%
  \pgfmathparse{int(#1)}%
  \ifnum\pgfmathresult>0
  \foreach \x in {0,...,#1}
     \filldraw (1.3*\x,-.1) circle (0.1);
  \foreach \x in {0,...,#1}
     \filldraw (1.3*\x,1.1) circle (0.1);
  \draw[line width=0.2mm] (-0.2,1.1) -- (8.0,1.1);
  \draw[line width=0.2mm] (-0.2,-.1) -- (8.0,-.1);
   \pgfmathtruncatemacro{\y}{#1}
  \foreach \x in {0,...,\y}
        \draw[line width=0.2mm] (1.3*\x,1.1) -- (1.3*\x,-.1);
  \foreach \x in {0,...,5} {
     \ifthenelse{\x=0 \OR \x=2 \OR \x=4}{
     \path [draw=blue,line width=\ra*0.0173986550528829mm,postaction={on each segment={mid arrow=blue}}]
        (1.3*\x+0.1,0) -- (1.3*\x+0.1,1) -- (1.3*\x+1.1,0)-- (1.3*\x+0.1,0);
      \draw (1.3*\x+0.4,0.35) node {+};
      \path [draw=red,line width=\ra*0.0173986550528829mm,postaction={on each segment={mid arrow=red}}]
         (1.3*\x+0.2,1.0) -- (1.3*\x+1.2,0.00)-- (1.3*\x+1.2,1.00) -- (1.3*\x+0.2,1.0);
      \draw (1.3*\x+0.85,0.65) node {-};
    } {
     \path [draw=red,line width=\ra*0.0173986550528829mm,postaction={on each segment={mid arrow=red}}]
        (1.3*\x+0.1,0) -- (1.3*\x+1.1,0) -- (1.3*\x+0.1,1)-- (1.3*\x+0.1,0);
      \draw (1.3*\x+0.4,0.35) node {-};
      \path [draw=blue,line width=\ra*0.0173986550528829mm,postaction={on each segment={mid arrow=blue}}]
         (1.3*\x+0.2,1.0) -- (1.3*\x+1.2,1.00) -- (1.3*\x+1.2,0.00) -- (1.3*\x+0.2,1.0);
      \draw (1.3*\x+0.85,0.65) node {+};
      }
    }
    \fi
  }

\def\chrldelta#1{%
  \pgfmathparse{int(#1)}%
  \ifnum\pgfmathresult>0
  \foreach \x in {0,...,#1}
     \filldraw (1.3*\x,-.1) circle (0.1);
  \foreach \x in {0,...,#1}
     \filldraw (1.3*\x,1.1) circle (0.1);
  \draw[line width=0.2mm] (-0.2,1.1) -- (8.0,1.1);
  \draw[line width=0.2mm] (-0.2,-.1) -- (8.0,-.1);
   \pgfmathtruncatemacro{\y}{#1}
  \foreach \x in {0,...,\y}
        \draw[line width=0.2mm] (1.3*\x,1.1) -- (1.3*\x,-.1);
  \foreach \x in {0,...,5} {
     \ifthenelse{\x=0 \OR \x=2 \OR \x=4}{
     \path [draw=red,line width=\ra*0.0254952100792158mm,postaction={on each segment={mid arrow=red}}]
        (1.3*\x+0.1,0) -- (1.3*\x+1.1,0) -- (1.3*\x+0.1,1)-- (1.3*\x+0.1,0);
      \draw (1.3*\x+0.4,0.35) node {-};
      \path [draw=purple,line width=\ra*0.0372949466940812mm,postaction={on each segment={mid arrow=purple}}]
         (1.3*\x+0.2,1.0) -- (1.3*\x+1.2,0.00) -- (1.3*\x+1.2,1.00) -- (1.3*\x+0.2,1.0);
      \draw (1.3*\x+0.85,0.65) node {-};
    } {
     \path [draw=purple,line width=\ra*0.0372949466940812mm,postaction={on each segment={mid arrow=purple}}]
        (1.3*\x+0.1,0) -- (1.3*\x+1.1,0) -- (1.3*\x+0.1,1)-- (1.3*\x+0.1,0);
      \draw (1.3*\x+0.4,0.35) node {-};
      \path [draw=red,line width=\ra*0.0254952100792158mm,postaction={on each segment={mid arrow=red}}]
         (1.3*\x+0.2,1.0) -- (1.3*\x+1.2,0.00) -- (1.3*\x+1.2,1.00) -- (1.3*\x+0.2,1.0);
      \draw (1.3*\x+0.85,0.65) node {-};
      }
    }
    \fi
  }

\def\kgusix#1{%
  \pgfmathparse{int(#1)}%
  \ifnum\pgfmathresult>0
  \foreach \x in {0,...,#1}
     \pgfmathtruncatemacro{\y}{2*\x+1}
     \draw (1.3*\x,-.1) node[circle,scale = 0.6, fill=black, black, draw, label=below:\y](b){};
  \foreach \x in {0,...,#1}
     \pgfmathtruncatemacro{\y}{2*\x+2}
     \draw (1.3*\x,1.1) node[circle,scale = 0.6, fill=black, black, draw, label=above:\y](b){};
  \draw[line width=0.2mm] (-0.2,1.1) -- (8.0,1.1);
  \draw[line width=0.2mm] (-0.2,-.1) -- (8.0,-.1);
   \pgfmathtruncatemacro{\y}{#1-1}
  \foreach \x in {0,...,\y} {
    \ifthenelse{\x=0 \OR \x=3 \OR \x=6}{
      \draw[line width=0.2mm] (1.3*\x,1.1) -- node[above] {$x'$} (1.3*\x+1.3,1.1);
      \draw[line width=0.2mm] (1.3*\x,-.1) -- node[below] {$x'$} (1.3*\x+1.3,-.1);
      } {
    \ifthenelse{\x=1 \OR \x=4 \OR \x=7}{
      \draw[line width=0.2mm] (1.3*\x,1.1) -- node[above] {$z'$} (1.3*\x+1.3,1.1);
      \draw[line width=0.2mm] (1.3*\x,-.1) -- node[below] {$z'$} (1.3*\x+1.3,-.1);
    }{
    \ifthenelse{\x=2 \OR \x=5 \OR \x=8}{
      \draw[line width=0.2mm] (1.3*\x,1.1) -- node[above] {$y'$} (1.3*\x+1.3,1.1);
      \draw[line width=0.2mm] (1.3*\x,-.1) -- node[below] {$y'$} (1.3*\x+1.3,-.1);
    }{}}
  }
  }
   \pgfmathtruncatemacro{\y}{#1}
  \foreach \x in {0,...,\y} {
    \ifthenelse{\x=0 \OR \x=3 \OR \x=6}{
      \draw[line width=0.2mm] (1.3*\x,1.1) -- node[left] {$z'$} (1.3*\x,-.1);
      } {
    \ifthenelse{\x=1 \OR \x=4 \OR \x=7}{
      \draw[line width=0.2mm] (1.3*\x,1.1) -- node[left] {$y'$} (1.3*\x,-.1);
    }{
    \ifthenelse{\x=2 \OR \x=5 \OR \x=8}{
      \draw[line width=0.2mm] (1.3*\x,1.1) -- node[left] {$x'$} (1.3*\x,-.1);
    }{}}
  }
  }
  \fi
  }

\def\wkgusixw#1{%
  \pgfmathparse{int(#1)}%
  \ifnum\pgfmathresult>0
  \foreach \x in {0,...,#1}
     \pgfmathtruncatemacro{\y}{2*\x+1}
     \draw (1.3*\x,-.1) node[circle,scale = 0.6, fill=black, black, draw, label=below:\y](b){};
  \foreach \x in {0,...,#1}
     \pgfmathtruncatemacro{\y}{2*\x+2}
     \draw (1.3*\x,1.1) node[circle,scale = 0.6, fill=black, black, draw, label=above:\y](b){};
  \draw[line width=0.2mm] (-0.2,1.1) -- (8.0,1.1);
  \draw[line width=0.2mm] (-0.2,-.1) -- (8.0,-.1);
   \pgfmathtruncatemacro{\y}{#1-1}
  \foreach \x in {0,...,\y} {
    \ifthenelse{\x=0 \OR \x=3 \OR \x=6}{
      \draw[line width=0.2mm] (1.3*\x,1.1) -- node[above] {$\Gamma_y\Gamma_z$} (1.3*\x+1.3,1.1);
      \draw[line width=0.2mm] (1.3*\x,-.1) -- node[below] {$\Gamma_y\Gamma_z$} (1.3*\x+1.3,-.1);
      \ifthenelse{\x=0}{
        \filldraw (1.3*\x+0.5,0.5) node (text) {$K^{k,l}_{i,j}$};
       \circledarrow{thick, black}{text}{0.37cm};
      }{
        \filldraw (1.3*\x+0.5,0.5) node (text) {$K^{k,l}_{i,j}$};
       \circledarrow{thick, black}{text}{0.37cm};
      }
      } {
    \ifthenelse{\x=1 \OR \x=4 \OR \x=7}{
      \draw[line width=0.2mm] (1.3*\x,1.1) -- node[above] {$K_z\Gamma_y$} (1.3*\x+1.3,1.1);
      \draw[line width=0.2mm] (1.3*\x,-.1) -- node[below] {$K_z\Gamma_y$} (1.3*\x+1.3,-.1);
      \filldraw (1.3*\x+0.5,0.5) node (text) {$\Gamma^{k,l}_{i,j}$};
      \circledarrow{thick, black}{text}{0.37cm};
    }{
    \ifthenelse{\x=2 \OR \x=5 \OR \x=8}{
      \draw[line width=0.2mm] (1.3*\x,1.1) -- node[above] {$K_y\Gamma_z$} (1.3*\x+1.3,1.1);
      \draw[line width=0.2mm] (1.3*\x,-.1) -- node[below] {$K_y\Gamma_z$} (1.3*\x+1.3,-.1);
      \filldraw (1.3*\x+0.5,0.5) node (text) {$\Gamma^{k,l}_{i,j}$};
      \circledarrow{thick, black}{text}{0.37cm};
    }{}}
  }
  }
   \pgfmathtruncatemacro{\y}{#1}
  \foreach \x in {0,...,\y} {
    \ifthenelse{\x=0 \OR \x=3 \OR \x=6}{
      \draw[line width=0.2mm] (1.3*\x,1.1) -- node[left] {$z'$} (1.3*\x,-.1);
      } {
    \ifthenelse{\x=1 \OR \x=4 \OR \x=7}{
      \draw[line width=0.2mm] (1.3*\x,1.1) -- node[left] {$y'$} (1.3*\x,-.1);
    }{
    \ifthenelse{\x=2 \OR \x=5 \OR \x=8}{
      \draw[line width=0.2mm] (1.3*\x,1.1) -- node[left] {$x'$} (1.3*\x,-.1);
    }{}}
  }
  }
  \fi
  }

\def\rungtripletA#1{%
  \pgfmathparse{int(#1)}%
  \ifnum\pgfmathresult>0
  \draw (-0.6,0.5) node (b){A};
  \foreach \x in {0,...,#1}
     \pgfmathtruncatemacro{\y}{2*\x+1}
     \draw (1.3*\x,-.1) node[circle,scale = 0.6, fill=gray, gray, draw](b){};
  \foreach \x in {0,...,#1}
     \pgfmathtruncatemacro{\y}{2*\x+2}
     \draw (1.3*\x,1.1) node[circle,scale = 0.6, fill=gray, gray, draw](b){};
  \draw[line width=0.2mm,gray] (-0.2,1.1) -- (8.0,1.1);
  \draw[line width=0.2mm,gray] (-0.2,-.1) -- (8.0,-.1);
   \pgfmathtruncatemacro{\y}{#1-1}
  \foreach \x in {0,...,\y} {
      \draw[line width=0.2mm,gray] (1.3*\x,1.1) -- (1.3*\x+1.3,1.1);
      \draw[line width=0.2mm,gray] (1.3*\x,-.1) -- (1.3*\x+1.3,-.1);
  }
  \pgfmathtruncatemacro{\y}{#1}
  \foreach \x in {0,...,\y} {
    \ifthenelse{\x=0 \OR \x=3 \OR \x=6}{
      \draw[line width=0.2mm,gray] (1.3*\x,1.1) --  (1.3*\x,-.1);
      } {
    \ifthenelse{\x=1 \OR \x=4 \OR \x=7}{
      \draw[line width=0.2mm,gray] (1.3*\x,1.1) -- (1.3*\x,-.1);
    }{
    \ifthenelse{\x=2 \OR \x=5 \OR \x=8}{
      \draw[line width=0.2mm,gray] (1.3*\x,1.1) --  (1.3*\x,-.1);
    }{}}
  }
  }
  \pgfmathtruncatemacro{\y}{#1}
  \foreach \x in {0,...,\y} {
    \ifthenelse{\x=0 \OR \x=1 \OR \x=6}{
        \filldraw (1.3*\x,0.5) node (text) {$t_x$};
        \filldraw[green, fill opacity=0.3] (1.3*\x,0.5) ellipse (0.24 and 0.9);
      }{}
    \ifthenelse{\x=2 \OR \x=3}{
        \filldraw (1.3*\x,0.5) node (text) {$t_y$};
        \filldraw[red, fill opacity=0.3] (1.3*\x,0.5) ellipse (0.24 and 0.9);
      }{}
    \ifthenelse{\x=4 \OR \x=5}{
        \filldraw (1.3*\x,0.5) node (text) {$t_z$};
        \filldraw[blue, fill opacity=0.3] (1.3*\x,0.5) ellipse (0.24 and 0.9);
      }{}
  }
  \fi
  }

\def\rungtripletB#1{%
  \pgfmathparse{int(#1)}%
  \ifnum\pgfmathresult>0
  \draw (-0.6,0.5) node (b){B};
  \foreach \x in {0,...,#1}
     \pgfmathtruncatemacro{\y}{2*\x+1}
     \draw (1.3*\x,-.1) node[circle,scale = 0.6, fill=gray, gray, draw](b){};
  \foreach \x in {0,...,#1}
     \pgfmathtruncatemacro{\y}{2*\x+2}
     \draw (1.3*\x,1.1) node[circle,scale = 0.6, fill=gray, gray, draw](b){};
  \draw[line width=0.2mm,gray] (-0.2,1.1) -- (8.0,1.1);
  \draw[line width=0.2mm,gray] (-0.2,-.1) -- (8.0,-.1);
   \pgfmathtruncatemacro{\y}{#1-1}
  \foreach \x in {0,...,\y} {
      \draw[line width=0.2mm,gray] (1.3*\x,1.1) -- (1.3*\x+1.3,1.1);
      \draw[line width=0.2mm,gray] (1.3*\x,-.1) -- (1.3*\x+1.3,-.1);
  }
  \pgfmathtruncatemacro{\y}{#1}
  \foreach \x in {0,...,\y} {
    \ifthenelse{\x=0 \OR \x=3 \OR \x=6}{
      \draw[line width=0.2mm,gray] (1.3*\x,1.1) --  (1.3*\x,-.1);
      } {
    \ifthenelse{\x=1 \OR \x=4 \OR \x=7}{
      \draw[line width=0.2mm,gray] (1.3*\x,1.1) -- (1.3*\x,-.1);
    }{
    \ifthenelse{\x=2 \OR \x=5 \OR \x=8}{
      \draw[line width=0.2mm,gray] (1.3*\x,1.1) --  (1.3*\x,-.1);
    }{}}
  }
  }
  \pgfmathtruncatemacro{\y}{#1}
  \foreach \x in {0,...,\y} {
    \ifthenelse{\x=0 \OR \x=5 \OR \x=6}{
        \filldraw (1.3*\x,0.5) node (text) {$t_y$};
        \filldraw[red, fill opacity=0.3] (1.3*\x,0.5) ellipse (0.24 and 0.9);
      }{}
    \ifthenelse{\x=3 \OR \x=4}{
        \filldraw (1.3*\x,0.5) node (text) {$t_x$};
        \filldraw[green, fill opacity=0.3] (1.3*\x,0.5) ellipse (0.24 and 0.9);
      }{}
    \ifthenelse{\x=1 \OR \x=2}{
        \filldraw (1.3*\x,0.5) node (text) {$t_z$};
        \filldraw[blue, fill opacity=0.3] (1.3*\x,0.5) ellipse (0.24 and 0.9);
      }{}
  }
  \fi
  }

\begin{document}

\title{The Heart of Entanglement: Chiral, Nematic, and Incommensurate Phases in the Kitaev-Gamma Ladder in a Field}
\author{Erik S. S{\o}rensen}
\email{sorensen@mcmaster.ca}
\affiliation{Department of Physics, McMaster University, Hamilton, Ontario L8S 4M1, Canada}
\author{Andrei Catuneanu}
\affiliation{Department of Physics, University of Toronto, Ontario M5S 1A7, Canada}
\author{Jacob Gordon}
\affiliation{Department of Physics, University of Toronto, Ontario M5S 1A7, Canada}
\author{Hae-Young Kee}
\email{hykee@physics.utoronto.ca}
\affiliation{Department of Physics, University of Toronto, Ontario M5S 1A7, Canada}
\affiliation{Canadian Institute for Advanced Research, CIFAR Program in Quantum Materials, Toronto, ON M5G 1M1, Canada}
\date{\today}

\begin{abstract}
The bond-dependent Kitaev model on the honeycomb lattice with anyonic excitations
has recently attracted considerable attention.
However, in solid state materials other spin interactions are present, and 
among such additional interactions, the off-diagonal symmetric Gamma interaction, another type of bond-dependent term, has been particularly
challenging to fully understand.
A minimal Kitaev-Gamma (KG) model 
has been investigated by various numerical techniques under a magnetic field, but definite conclusions about field-induced spin liquids
remain elusive and one reason may lie in the limited sizes of the two-dimensional geometry it is possible to access numerically.
We therefore focus on the KG model defined on a two-leg ladder which is much more amenable to a complete study,
and determine the entire phase diagram in the presence of a magnetic field along $[111]$-direction.
Due to the competition between the interactions and the field, an extremely rich phase diagram emerges with fifteen distinct phases.
Focusing near the antiferromagnetic Kitaev region, we  identify nine different phases solely within this region:
several incommensurate magnetically ordered phases, spin nematic, and {\it two} chiral phases with enhanced entanglement.
Of particular interest is a  highly entangled phase with {\it staggered chirality} with zero-net flux occurring at intermediate field, which
along with its companion phases outline a heart-shaped region of high entanglement, the heart of entanglement.
We compare our results for the ladder with a C$_3$ symmetric cluster of the two-dimensional honeycomb lattice, and offer 
insights into possible spin liquids in the two-dimensional limit.
\end{abstract}
\maketitle

\section{Introduction}
The Kitaev model on a two-dimensional honeycomb lattice is a rare example of an
exactly solvable model offering a quantum spin liquid with fractional excitations.\cite{kitaev2006}
Under a time-reversal symmetry breaking field, it exhibits non-Abelian anyons 
with half-quantized thermal Hall conductivity originated from Majorana edge mode.
Since the original proposal, finding a solid state material possessing such a quantum spin liquid has attracted great attention. 
A microscopic mechanism for realizing the Kitaev model in solid-state material was first suggested using the combined effects of
strong spin-orbit coupling and electron-electron interactions\cite{khaliullin2005orbital,jk2009prl}.
Later, the nearest neighbor generic spin model on an ideal honeycomb lattice was re-derived,
and it was found that there are additional bond-dependent interactions present with the so-called Gamma ($\Gamma$) interaction among the most intriguing.\cite{rau2014prl} 

From the material perspective, $\alpha$-RuCl$_3$ was proposed as a leading candidate with a weaker coupling between layers
making the material close to two dimensional (2D).\cite{plumb2014prb,HSKim2015prb,koitzsch2016prl,sandilands2016spinorbit,zhou2016arpes,banerjee2016proximate} 
Furthermore, the $\Gamma$ interaction has been found to be as large as the Kitaev interaction in $\alpha$-RuCl$_3$.\cite{HSKim2016structure,janssen2017model,winter2016challenges}
Since then, RuCl$_3$ has been explored by several experimental and theoretical techniques.\cite{rau2016review,winter2017review,hermanns2018review,takagi2019review,janssen2019review}
In particular, early inelastic neutron scattering\cite{banerjee2016proximate} and Raman spectroscopy\cite{sandilands2015continuum} 
measurements have suggested a strong frustration well above the magnetic ordering
temperature, indicating strong frustration which may originate from the bond-dependent Kitaev and $\Gamma$ interactions. 
Remarkably, a half-quantized thermal Hall conductivity was recently reported in $\alpha$-RuCl$_3$ in a certain range of the magnetic field
\cite{kasahara2018thermal}
when a zig-zag magnetically ordered phase is destroyed by the magnetic field.\cite{johnson2015monoclinic}
Other physical quantities accessed by several experimental techniques in RuCl$_3$ also 
suggested that there is a nontrivial intermediate phase under a magnetic field which is different from
a trivially polarized spin state in high field regime.\cite{baek2017evidence,zheng2017gapless,leahy2017thermal,hentrich2018phonon,lampenkelley2018induced,jansa2018types,banerjee2018excitations,Shi2018induced,widmann2019thermodynamic,Saha2020highfield}  
However, the experimental evidence for a 
 field-induced intermediate disordered phase in RuCl$_3$ is still under debate.\cite{wolter2017induced,yu2018prl,yamashita2020sample}
 
In parallel to the experimental progress, theoretical attempts to find nontrivial field-induced phases in 
extended Kitaev model have been pursued extensively.
 \cite{yadav2016field,janssen2017model,chern2017KH-field,fu2018robust,Liang2018response,gohlke2018dynamical,nasu2018successive,
 Ronquillo2019signatures,liang2018intermediate,liu2018dirac,Kaib2019field,hickey2019visons,Chern2020Magnetic,Lee2020Magnetic,gohlke2020arXiv}
Most numerical studies are limited to either near the antiferromagnetic (AFM) Kitaev or near the ferromagnetic (FM) Kitaev region, 
as the exactly solvable Kitaev point offers a starting point.
In particular, a minimal Kitaev-Gamma (KG) model 
under a $[111]$ magnetic field has been studied near FM Kitaev and AFM $\Gamma$ region relevant to RuCl$_3$.\cite{Gordon2019,Chern2020Magnetic,Lee2020Magnetic,gohlke2020arXiv}
A 24-site exact diagonalization (ED) study showed a field-revealed Kitaev spin liquid near FM Kitaev region with a finite AFM
 $\Gamma$ interaction,
when the magnetic field is tilted away from the $[111]$-axis.\cite{Gordon2019} However, the infinite tensor product state (iTPS)
 found a small confined Kitaev spin liquid, and broken C$_3$ rotational phases 
are induced under the magnetic field.\cite{Lee2020Magnetic}
Interestingly various large unit cell magnetic orderings 
have been reported in the classical KG model under the magnetic field along [111]-axis.\cite{Chern2020Magnetic}, which are replaced by these broken rotational phases in quantum model.\cite{gohlke2020arXiv}
Whether the C$_3$ broken phases are quantum spin liquids or not is at present not clear and will require further studies.

Numerical studies near AFM Kitaev limit under a magnetic field have found intriguing results.\cite{fu2018robust,gohlke2018dynamical,hickey2019visons,lu2018spinon,zou2020neutral,Ronquillo2019signatures,liang2018intermediate,patel2019spinon,nasu2018successive}
It was suggested that
a gapless U(1) spin liquid is induced by the magnetic field.\cite{hickey2019visons}
The energy spectra obtained by 24-site ED showed putative gapless excitations in the intermediate field,
which then transition to a polarized state (PS) in high field regime.
Several iDMRG studies also reported a change of central charge depending on the number of legs in the DMRG which indicates
a finite spinon Fermi surface.\cite{lu2018spinon}  However  one may question if the dense energy spectra are due to incommensurate order
which is difficult to detect due to the finite size of ED and limited access to momentum points in iDMRG. 
Indeed, different iDMRG studies have found different gapless points in momentum space.\cite{lu2018spinon,gohlke2018dynamical,patel2019spinon} 


Despite intensive studies, definite conclusions on possible phases and the nature of numerically determined phases near the Kitaev regions remain indefinite.
One reason for the controversial results among the previous studies may lie in the limited sizes of the two-dimensional honeycomb geometry that one can access numerically. 
Furthermore, the zero-field and field-induced phases of the entire phase space of KG model are yet to be determined. 
We therefore focus on the KG model defined on a two-leg ladder which is much more amenable to a detailed study 
and a complete phase diagram in the presence of a magnetic field along $[111]$-direction can be determined.
%
Using high through-put iDMRG calculations we map out the entire phase diagram of the KG ladder which shows an extremely rich structure.
After we determine the entire phase diagram, we focus on the region of the phase diagram where the Kitaev interaction is predominantly AFM. 
In this region, in the absence of a magnetic field, we identify a novel spin nematic (SN) phase with quadropolar order in addition to the AFM Kitaev phase (\AK) and a phase connected
to the isotropic FM ladder through a local 6-site spin rotation, FM$_{U_6}$.\cite{chaloupka2015hidden,kimchi2016duality,perkins2018magnetic,stavropoulos2018counter}
In zero field the KG ladder can be mapped to a ladder with four-spin exchange closely related to JQ models~\cite{Sandvik2007} extensively studied as models of deconfined criticality~\cite{Senthil2004}.

When a magnetic field in the $[111]$ direction is introduced, field and spin interactions compete, and a proliferation of phases is observed.
We identify phases with scalar chiral ordering and several phases with magnetic ordering some of which might display incommensurate or very large unit cell
ordering.  Of particular interest is two chiral ordered phases characterized by a staggered chirality (SC) and uniform chirality (UC).
The SC phase is a magnetically disordered and highly entangled phase occurring at intermediate magnetic field above the \AK\ phase.
It has the staggered chirality with zero-net flux despite it is under a rather large external field, and
shows clear chain end excitations. Rather poetically, this phase along with its companion phases outline a heart-shaped region of high entanglement, `the chiral heart of entanglement'.
On the other hand, the UC phase with the uniform chirality leading to a finite net flux appears between the SN and 
and the rung singlet (RS$_{U_6}$) phase connected
to the isotropic AFM ladder through a local 6-site spin rotation.\cite{chaloupka2015hidden,kimchi2016duality,perkins2018magnetic,stavropoulos2018counter}
It emerges at extremely low field, as if three phases, SN, UC, and RS$_{U_6}$  may meet at a critical point. 
All together, near the AFM Kitaev region alone, we identify 9 possible distinct phases in addition to the FM$_{U_6}$ phase
and the PS occurring at high magnetic fields.

Below, in section~\ref{sec:model},  we first review the KG model. In addition to the extensive 2D cluster studies,
a one-dimensional (1D) KG chain model  including only $x$- and $y$-bonds was studied 
using non-Abelian bosonization and DMRG and reported SU(2) emergent phases.\cite{Wang2020a}.
The two-leg ladder is made of two such KG chains by connecting them by $z$-bond. 
Technical details relevant fo the iDMRG and DMRG numerical methods are subsequently discussed in the following section, ~\ref{sec:num}. 
An overview and detailed discussion of the full phase diagram of the ladder is presented in section~\ref{sec:full} along with a discussion of 
the connection between the KG chain and the ladder.
In section~\ref{sec:afm} we focus on the vicinity of the AFM  Kitaev region under a magnetic field, where a rich phase diagram with various phases with enhanced entanglement
is found.  
Finally, in section~\ref{sec:compare} 
we compare our results for the ladder with 24-site ED results obtained in the honeycomb geometry, and discuss 
the implications of the ladder results to the 2D limit.


\section{The Two-leg Kitaev-Gamma (KG) ladder}\label{sec:model}
\begin{figure}
  \centering
  \includegraphics[width=0.5\textwidth]{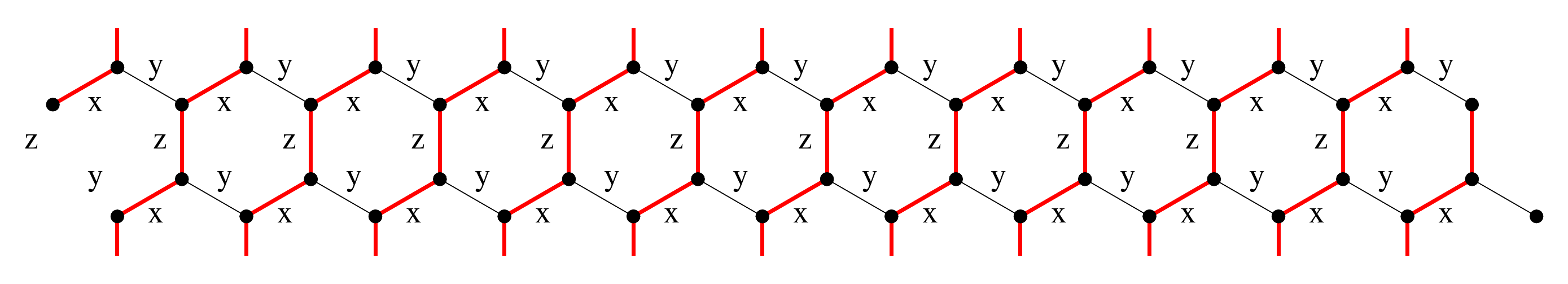}
   \caption{Two-leg ladder KG honeycomb strip with alternating $x$ and $y$ bonds along the leg and $z$-bond between the chains.
 }
  \label{fig:ladder}
\end{figure}
The two-leg KG ladder is formed out of a strip of the honeycomb lattice. The two KG chains with only $x$- and $y$-bonds are coupled by adding the $z$-bond as shown in Fig. \ref{fig:ladder}.
Periodic boundary conditions in the direction perpendicular to the ladder is then imposed by directly coupling the dangling $z$-bonds thereby forming a regular ladder. 
Sometimes, the additional $z$-bonds from imposing periodic boundary conditions are taken to be of opposite sign (and or strength) in which case the 
resulting model is usually referred to as a honeycomb ladder~\cite{Kivelson2013}.
Here, all z-bonds are identical and a regular ladder is formed.
In addition to the bond-dependent Kitaev interaction,
the KG Hamiltonian incorporates another bond-dependent interaction, $\Gamma$.\cite{rau2014prl}
For the KG ladder we orient the bonds so that
the Kitaev $z$-bond connects the two legs of the ladder as shown in Fig.~\ref{fig:ladder}. The complete Hamiltonian is then given by
\begin{equation}
   H_{KG} = \sum_{\langle i,j\rangle_{\gamma \in (x,y,z)}}  K S_i^\gamma S_j^\gamma + \Gamma\left(S^\alpha_iS^\beta_j + S^\beta_iS^\alpha_j\right) 
   \label{eq:KGH}
\end{equation} 
where $(\alpha, \beta)$ takes on the values $(y,z)/(x,z)/(x,y)$ for $\gamma = x/y/z$, and $\langle i, j \rangle$ refers to the nearest neighbor sites.
We keep $K = \cos\phi$ and $\Gamma=\sin\phi$ and interpolate between the Kitaev ladder and $\Gamma$ ladder by varying $\phi$ from 0 to $2\pi$.
We denote the total number of sites in the ladder (including both legs) by $N$.
%
The pure Kitaev ladder at $\phi = 0, \pi$ is exactly solvable~\cite{feng2007characterization} and at both points
it is in a disordered gapped phase\cite{feng2007characterization}. However, recently it was shown that a non-local string order parameter (SOP) 
can be defined~\cite{Catuneanu2018ladder} at the Kitaev points. The SOP remains non-zero in the presence of a small Heisenberg coupling, $J$, at both
the Kitaev points. Close to the FM Kitaev point, $\phi=\pi$, the phase diagram of the KG ladder has recently been investigated in the presence of a magnetic
field and additional interactions~\cite{Gordon2019}.  On the other hand, relatively little is known about the rest of the phase diagram of the KG ladder which is our focus here.

       In one dimension in the absence of a magnetic field, the closely related KG chain has been investigated in considerable detail~\cite{Wang2020a,Wang2020b}. An extended
disordered phase close to the AFM Kitaev point, $\phi=0$ has been identified along with an adjacent spin nematic phase. For the KG chain it can rigorously be established that
the phase diagram is symmetric with respect to $\Gamma \to -\Gamma$, a symmetry that is clearly absent in the KG ladder. 

It is of particular interest to also consider the effect of a magnetic field. Here we exclusively consider a field in the $[111]$ direction, perpendicular to the honeycomb plane of the ladder.
The magnetic field leads to a Zeeman coupling as 
\begin{equation}
H = H_{KG} - g \frac{\mu_B}{\hbar} \sum_i {\bf h} \cdot {\bf S},
\end{equation}
where we choose the direction of ${\bf h}$ along the [111]-axis normalized 
as ${\bf h} = \frac{h}{\sqrt{3}} (1,1,1) $, $g=1$, and ${\bf S} = \hbar \frac{\sigma}{2}$ with $\sigma$ is Pauli matrix. 
We use units with $\hbar=1$ and $\mu_B=1$.

\section{Numerical methods}\label{sec:num}
As our main tools for investigating the KG ladder we use exact diagonalization (ED), finite size density matrix renormalization group (DMRG) and infinite DMRG (iDMRG) techniques.
The iDMRG calculations are performed in two different ways. A high through-put mode with a small unit cell of size 24 or 60 and a maximal bond dimension of 500 and a high precision mode
with a unit cell of 60 and a maximal bond dimension of 1000. Typical precisions for the two iDMRG modes are $\epsilon=10^{-8}$ and $\epsilon=10^{-10}$, respectively, and $\epsilon=10^{-10}$
for the finite size DMRG calculations. Finite size DMRG calculations are performed both with open boundary conditions (OBC) and for smaller system sizes with periodic boundary conditions (PBC).
        In order to establish the phase diagram we focus on several different characteristics. With $e_0$ the ground-state energy per spin we define the energy susceptibilities
\begin{equation}
  \chi_h^e = -\frac{\partial^2 e_0}{\partial h^2},\ \ \chi_\phi^e = -\frac{\partial^2 e_0}{\partial \phi^2},
\end{equation}
where $\chi_h^e$ could equally well be called a magnetic susceptibility.
For finite size systems of size $N$ it has been established~\cite{Albuquerque2010} that the energy susceptibility at a quantum critical point (QCP) diverges as
\begin{equation}
  \chi^e \sim N^{2/\nu-d-z}.
\end{equation}
Here $\nu$ and $z$ are the correlation and dynamical critical exponents and $d$ is the dimension.
It follows that $\chi^e$ may not necessarily detect the phase transition if the critical exponent $\nu$ is sufficiently large. We have therefore found it useful
to supplement the analysis of $\chi^e$ by a study of the entanglement spectrum~\cite{Li2008} (ES) at different sections of the ladder. We have found it most useful to
use a partition of the ladder where a cut is introduced at site $N/2-1$ thereby intersecting a rung. With $\lambda_\alpha$ the eigenvalues of the resulting reduced density
matrix, $\rho_{N/2-1}$ the entanglement spectrum can be defined as $-\ln\lambda_\alpha$ and yields a characteristic signature of a phase. Close to the QCP the $\lambda_\alpha$ rapidly
change whereas  they remain approximately constant inside a phase. We therefore focus on the largest of the $\lambda_\alpha$'s which we denote by $\lambda_1$ and study
the lowest edge of the entanglement spectrum defined by
\begin{equation}
  -\ln\lambda_1.
\end{equation}
In a product phase $\lambda_1=1$ ($-\ln\lambda_1=0$) and such phases, with zero entanglement, are therefore easily detected by tracing out $-\ln\lambda_1$. Conversely,
phases with high entanglement will have $-\ln\lambda_1\gg 0$ and of course, due to the ordering of the $\lambda_\alpha$ one must have 
$-\ln\lambda_\alpha>-\ln\lambda_1$ for any $\alpha>1$. 
  Sometimes changes in $-\ln\lambda_1$ are imperceptible and we have therefore found it useful to define an entanglement spectrum susceptibility as follows
\begin{equation}
  \chi_h^{\lambda_1} = -\frac{\partial^2 \lambda_1}{\partial h^2},\ \ \chi_\phi^{\lambda_1} = -\frac{\partial^2 \lambda_1}{\partial \phi^2}.
\end{equation}
Since the ES {\it has} to change at the QCP, $\chi^{\lambda_1}$ should be able to detect {\it any} phase transition with the exception of unlikely scenario's where $\lambda_1$ only changes
linearly at the QCP with all non-trivial changes in the higher $\lambda_\alpha$'s. We have found $\chi^{\lambda_1}$ to be an {\it extremely} sensitive measure, often changing many orders
of magnitude at a QCP, and we therefore typically focus on $\ln\chi^{\lambda_1}$.

\begin{figure}[t!]
  \centering
 \includegraphics[width=0.5\textwidth]{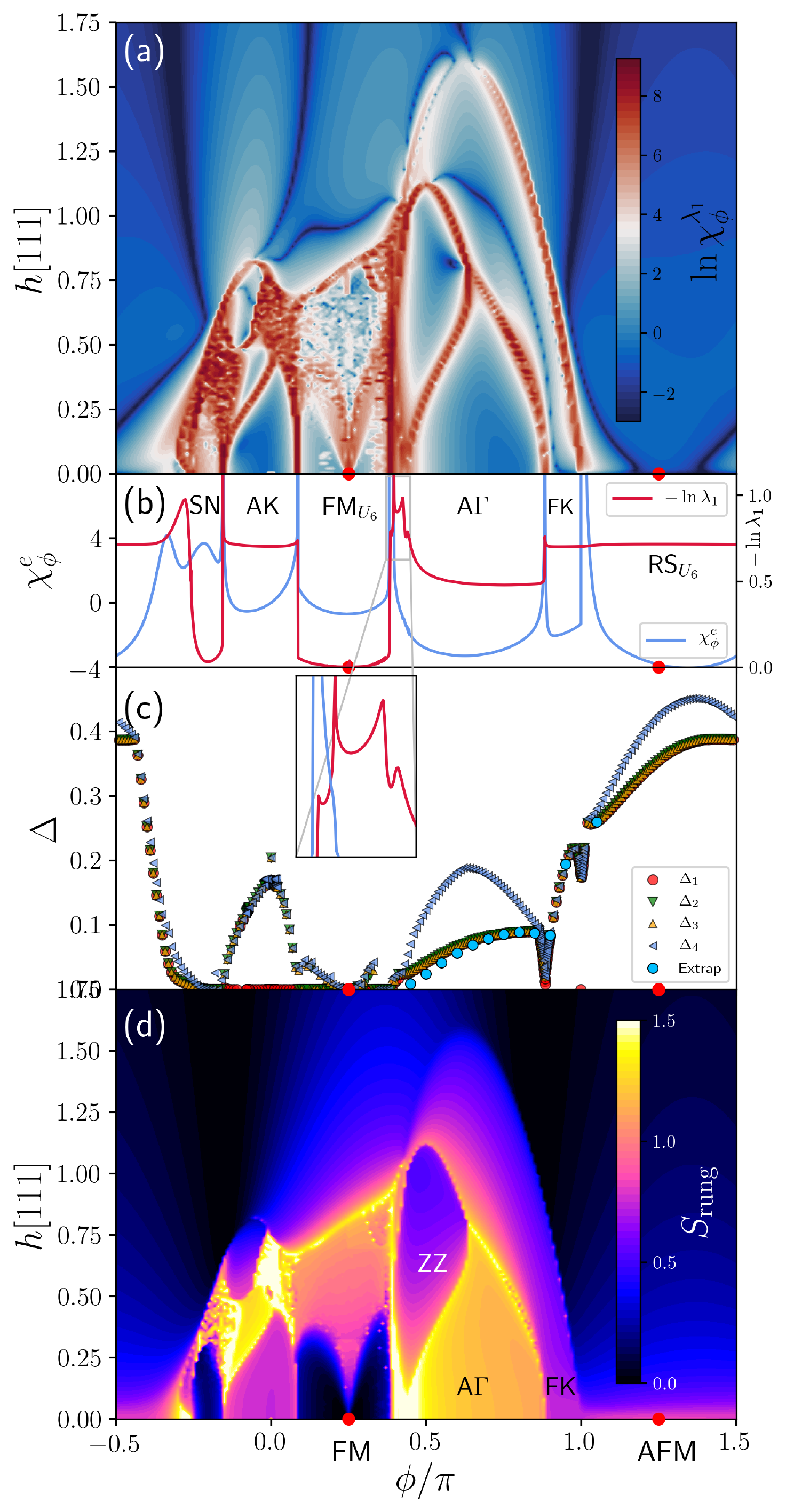}
  \caption{Phase diagram of the two-leg KG ladder model versus field, $h[111]$ and coupling $\phi/\pi$. 
  {\bf (a)} The entanglement spectrum susceptibility $\chi_\phi^{\lambda_1}$ on a logarithmic scale.
  {\bf (b)} The energy susceptibility, 
  $\chi_\phi^e$ (blue line) versus $\phi$ in zero field, and the lowest edge of the entanglement spectrum, $-\ln\lambda_1$ (red line), versus $\phi$ in zero field. 
  {\bf (c)} The excitation gap to the first 4 states, solid blue circles are extrapolations of $\Delta_1$ to $N=\infty$.
        {\bf (d)} $S_{\rm rung}$ versus $\phi$ and $h[111]$. 
        See labels in Fig.~\ref{fig:FineAFK}.
        Results in {\bf (a)} and {\bf (d)} are from high through-put iDMRG calculations with a unit cell of 60 sites with $\Delta\phi=0.01\pi,\ \Delta h[111]=0.01$
        Results in {\bf (b)} are from high precision iDMRG calculations with a unit cell of 60 sites with $\Delta\phi=0.001\pi$. 
        Results in {\bf (c)} are from high precision finite size DMRG calculations with PBC, $N=60$ and $\Delta\phi=0.01\pi$. 
        Close to $\phi=\pi$ and $\phi=0.88\pi$, $\Delta\phi=0.001\pi$.
  }
 \label{fig:FullPD}
\end{figure}
\section{Full KG ladder $\phi,h[111]$ phase diagram}\label{sec:full}
We start with a discussion of the full phase diagram covering the entire range $\phi\in [0,2\pi]$ and $h[111]\in [0,1.75]$. Our results for the full phase diagram as obtained from iDMRG calculations
are shown in Fig.~\ref{fig:FullPD}. We first show $\ln\chi^{\lambda_1}_\phi$ in
Fig.~\ref{fig:FullPD}(a). The divergence of $\chi^{\lambda_1}_\phi$ at a phase transition is so strong that it is most sensible to plot $\ln\chi^{\lambda_1}_\phi$. A well defined phase, where $\lambda_1$ is close to constant
is then visible in Fig.~\ref{fig:FullPD}(a) as a dark blue coloring. On the other hand, a divergent $\chi^{\lambda_1}_\phi$, indicating a phase transition is visible as a dark red color. As is clearly evident from Fig.~\ref{fig:FullPD}(a)
the complexity of the phase diagram due to the many competing phases is truly remarkable. 

Fig. 2(b) and (c) show $\chi^e_\phi$ and $- \ln \lambda_1$, and the spin excitation gap at zero field, respectively.
We will discuss them in detail later.

A second view of the full phase diagram is shown in Fig.~\ref{fig:FullPD}(d) where the bipartite von Neumann entanglement entropy $S_\mathrm{rung}$ is shown. We define
\begin{equation}
  S_\mathrm{rung} = -\mathrm{Tr} \rho_{N/2-1}\ln\rho_{N/2-1},
\end{equation}
where $\rho_{N/2-1}$ is the reduced density matrix obtained from a partition of the ladder after site $N/2-1$, a partition that will cut a {\it rung} in the ladder. Highly entangled phases are here visible
as bright yellow colors where as phases with negligible or no entanglement, where the ground-state wave function is well described by a simple product form, are shown as dark blue.

We note that considerable scattering are clearly visible in certain regions at finite fields, for instance above $\phi=\pi/4$ and $\phi=0$. The noise is due to poor convergence of the iDMRG due
to the high frustration present. In these regions a more careful analysis with either exact diagonalizaiton or finite-size DMRG is necessary.

\subsection{Zero field phase diagram}
Let us now focus on the phase diagram in zero field. A detailed high precision calculation of $\chi_\phi^e$ at zero field is shown in Fig.~\ref{fig:FullPD}(b) along with the lower edge of
the entanglement spectrum (ES), $-\ln\lambda_1$. A large number of well defined phase transitions are clearly visible which we now discuss.

Part of the phase diagram close to the FM Kitaev point $\phi=\pi$, 
has previously been discussed~\cite{Gordon2019} and in zero field a gapped spin liquid phase denoted KSL between \tcRStoFK\ and \tcFKtoAG\ along with a second gapped phase K$\Gamma$SL starting below \tcFKtoAG\ have been identified. 
Since we here discuss the full phase diagram we shall refer to the KSL phase as \FK\ and the K$\Gamma$SL phase as \AG\ to distinguish them from the phases ocurring at the AFM Kitaev point. 
The notation \AG\ makes sense since this phase surrounds $\phi=\pi/2$ where $K=0$, $\Gamma=1$.

A {\it local}
unitary transformation, $U_6$, is also known~\cite{CatuneanuThesis2019,Wang2020a,Wang2020b}. The $U_6$ transformation locally rotates the spins in a manner so that the $\Gamma$ couplings are transformed
into Heisenberg like ($xx$, $yy$, or $zz$) couplings with a changed sign.  The transformation can be applied equally well to the chain, the ladder and the honeycomb plane. At the points
$\phi=\pi/4$ and $\phi=5\pi/4$ where the Kitaev and $\Gamma$ couplings are of equal strength the KG ladder is therefore transformed into an isotropic FM and AFM
Heisenberg ladder. These two points therefore have hidden SU(2) symmetry. It is well established that the isotropic AFM Heisenberg ladder is in a gapped disordered rung singlet (RS) phase~\cite{Dagotto1992,Barnes1993} and we therefore denote the corresponding phase for the KG ladder
as RS$_{U_6}$. 
In the two-dimensional honeycomb lattice limit the RS$_{U_6}$ phase becomes a 120 degree ordered phase.\cite{rau2014prl}  
For the FM point, $\phi=\pi/4$ we denote the magnetically ordered gapless phase by FM$_{U_6}$. Since the FM$_{U_6}$ is well approximated by a product wave function with negligible entanglement it
is distinctly visible in Fig.~\ref{fig:FullPD}(d) with its almost black coloring.
The FM and AFM points are shown as solid red circles along the $\phi$-axis in Fig.~\ref{fig:FullPD}.
The same unitary $U_6$ transformation transforms the FM Kitaev point, $\phi=\pi$, to the AFM Kitaev point, $\phi=0$. The energy spectrum at these two points must therefore be identical,
a property that does not hold for any non-zero $\Gamma$.

Starting from right to left we observe
the transition from RS$_{U_6}$ to the \FK\ phase occurs at \tcRStoFK\ with the subsequent transition from the \FK\ phase to the \AG\ phase occurring at \tcFKtoAG. Between the gapless FM$_{U_6}$ phase
and the gapped \AG\ phase we observe a rapid sequence of several well defined phase transitions at \tcAGtoSNP, \tcSNPtoQ, \tcQtoQP\ and finally at \tcQPtoFM. 
See the zoomed inset in Fig.~\ref{fig:FullPD}(b). While $\chi_\phi^e$ only detects a single transition at \tcQPtoFM\ the other 3 transitions are clearly identifiable in $-\ln\lambda_1$ as shown in 
the zoomed inset.
The precise nature of the intervening phase
is at present unclear and left for future study. A clear transition out of the FM$_{U_6}$ phase to the gapped spin liquid phase \AK\ is observed at \tcFMtoAK. As previously mentioned,
precisely at $\phi=0$ a non-local string order has been found~\cite{Catuneanu2018ladder} and the \AK\ phase is clearly identifiable as a spin liquid phase. In the 2D honeycomb lattice
limit the \AK\ phase becomes the AFM Kitaev spin liquid. We shall discuss the \AK\ phase further below.
We now turn to a discussion of the last phase observed in zero field, to the left of the \AK\ phase.

\begin{figure}[tbh]
  \centering
        \includegraphics[width=0.5\textwidth]{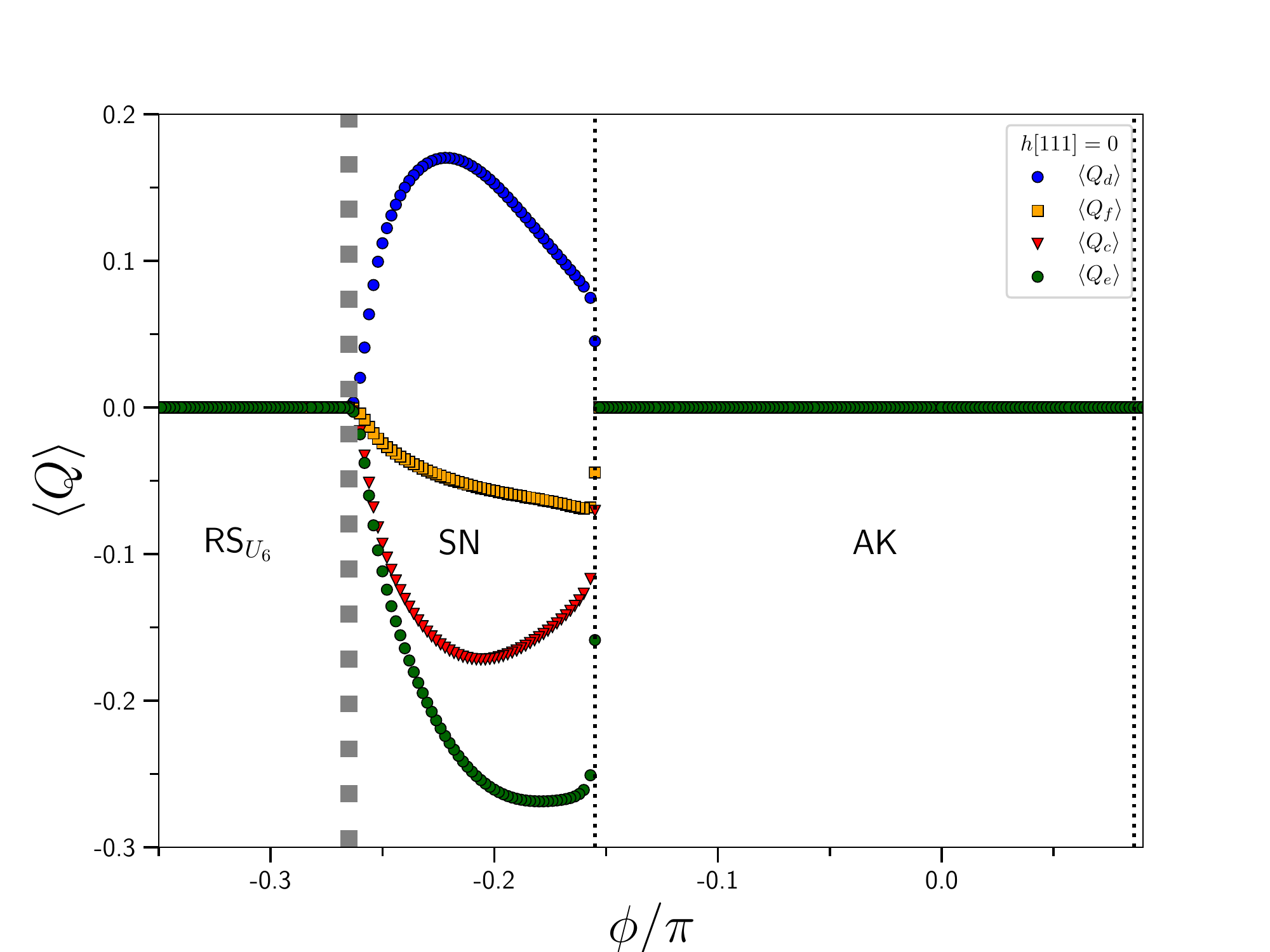}
  \caption{Nematic ordering at zero field in the AFM Kitaev region. The nematic order parameters are determined along {\it one} leg of the KG ladder.
  Results are from high precision iDMRG calculations with a unit cell of 60. 
  The dotted vertical lines denote the transitions at critical fields determined from divergences in $\chi_\phi^e$ and $\chi_\phi^{\lambda_1}$.
  }
  \label{fig:Nematic}
\end{figure}

\subsubsection{Nematic Phase in zero Field, SN}
In a recent study~\cite{Wang2020b} the KG chain was investigated and a phase with spin-nematic (spin-quadropole) order adjacent to the spin-liquid phase at $\phi=0$ was identified.
From a symmetry analysis of the following 4 order parameters were identified:
\begin{eqnarray}
  Q_c &=& S^x_{j}S^y_{j+1}+S^y_jS^x_{j+1}\nonumber\\
  Q_d &=& S^x_{2n+1}S^z_{2n+2}+S^z_{2n+1}S^x_{2n+2}+\nonumber\\
      &\ &    S^y_{2n+2}S^z_{2n+3}+S^z_{2n+2}S^y_{2n+3},\ \ n=0,1,\ldots \nonumber\\
  Q_e &=& S^y_{2n+1}S^y_{2n+2}+S^x_{2n+2}S^x_{2n+3},\ \ n=0,1,\ldots\nonumber\\
  Q_f &=& S^z_{j}S^z_{j+1}\nonumber\\
\end{eqnarray}
The two order parameters $Q_c$ and $Q_d$ describe off-diagonal ordering between sites not coupled by the same terms in the Hamiltonian and $Q_e$, $Q_f$ diagonal ordering
again between sites not coupled the same way in the Hamiltonian. The above definitions therefore depend on a specific ordering of the couplings along the leg. Considering
only the Kitaev coupling, site 1,2 would be $xx$ coupled, site 2,3 $yy$ coupled and so forth. (Note that along the legs of the KG ladder no $S^zS^z$ coupling occurs so $Q_f$ can be defined between any two nearest neighbor sites). 
We can use the same order parameters to study nematic ordering along the legs of the KG ladder. Our results are shown in Fig.~\ref{fig:Nematic} as obtained from iDMRG. Due to the translational
invariance the $Q$'s are the same among all sites and are easily calculated. They all four become non-zero at \tcAKtoSN\ which coincides with divergences in $\chi_\phi^e$ and $\chi_\phi^{\lambda_1}$.
%
%
The transition by varying $\phi$ is also clearly visible directly in $-\ln\lambda_1$ as can be seen in Fig.~\ref{fig:FullPD}(b). 

However, this is not a conventional spin-quadropole phase.
The DRMG with OBC shows two different magnetic orderings depending on the size of the system.
One has the AFM order along the leg and FM between the rung, 
and the other has a 6-site ordering mapping to AFM order after 6-site transformation, while the iDMRG finds the first one.
The AFM order corresponds to a stripe order, if we continue the ordering pattern by increasing the number of  ladders to the the 2D limit.
Let us briefly consider what happens if an additional Heisenberg coupling $J$ is introduced along side the $K$ and $\Gamma$ terms.
In that case, the striped phase appears for an AF Heisenberg interaction $J>0$ in 24-site ED calculations on the C$_3$ cluster, while the second ordering is likely a spiral order occurring for $J <0$.\cite{rau2014prl}
This suggests that this particular window of $\phi$ with $J=0$ is in fact a line of first order transitions (in $J$) between these two orderings.
To confirm such a possibility, we have studied the phase boundary by sweeping $J$ ( parameterized as $J\equiv K \cos\theta )$. 
Indeed we find a clear first order transition occuring at $J=0$
in $\chi_\theta^e$, the second derivative of ground state energy per spin with respect to $\theta$, as shown Appendix~\ref{app:SN}.

While it is a line of first order transitions, we would refer to this as SN for spin-nematic, as the SN is a common feature of these orderings coexiting along
the transition line. As the magnetic field becomes finite, it develops a magnetic order with almost zero entanglement which is shown later.
The transition
out of the SN phase into the RS$_{U_6}$ phase occurs at around \tcSNtoRS. As we shall discuss later, this quantum phase transition is actually a 
multi-critical point where the first order transition line ends, and two other phases occur. A high
precision determination of the location of the critical point is therefore significantly more difficult than for the other QCP's. In particular so, since the entanglement
for $\phi<$\tcSNtoRS\ is exceedingly high. It is therefore shown as a broader dashed line in Fig.~\ref{fig:Nematic}.


\subsubsection{Excitation gap in Zero field}
In Fig.~\ref{fig:FullPD}(c) we show the spin excitation gap to the first 4 lowest lying states, $\Delta_1$, $\Delta_2$, $\Delta_3$ and $\Delta_4$ as obtained
from high precision finite size DMRG on ladders with $N=60$. We have verified that finite size effects are relatively small if not in the proximity of a QCP. This is indicated
by the solid blue circles in Fig.~\ref{fig:FullPD}(c) which indicate extrapolations to $N=\infty$ of $\delta_1$ by fitting data for $N=24,36,48,60$ and $72$ to the form $\Delta(L)=\Delta_\infty+a \exp(-L/\xi)/L$.

Starting from the right, we find that the RS$_{U_6}$ phase, as expected, has a single ground-state with
a well defined triplet excitation through-out most of the phase. The triplet excitation merges with higher lying excitations at $\phi=1.56\pi$. Precisely
at the FM Kitaev point, $\phi=\pi$, 
there is a level crossing leading to a clear first order transition.
This is exact also for finite systems. The same holds true by symmetry at the AFM Kitaev point $\phi=0$. 
However, 
while the \FK\ phase has a single ground-state below a well defined gap and $\phi=\pi$ is a transition point, 
in the \AK\ phase the ground-state remains
double degenerate below a gap, only split by finite-size effects. 
The \AG\ phase also has a single ground state with a well defined triplet excitations above it. 
The FM$_{U_6}$ phase is gapless, but occasionally excited state DMRG calculations get trapped in higher lying states and the gapless nature
of the phase does not appear clearly in Fig.~\ref{fig:FullPD}(c). The final phase we discuss here, the SN phase occurring between \tcSNtoRS\ and \tcAKtoSN\
is clearly gapless as can be seen in Fig.~\ref{fig:FullPD}(c) with all four gaps close to zero.

\subsection{Relation to 1D KG Chain}
As eluded to above, the phase diagram of the ladder is closely connected to that of the KG chain and the KG model defined on the full two-dimensional
honeycomb lattice.

Comparing to the phase diagram of the KG chain~\cite{Wang2020a,Wang2020b}, a gapped \FK\ phase appeared in the ladder, while it was gapless in the chain model. Similarly, the \AK\ persists near AFM Kitaev region,
and it is gapped in the ladder, while gapless in the chain. 
RS$_{U_6}$ is gapped in the ladder similar to AFM Heisenberg model, where as it was gapless in the chain. FM$_{U_6}$ remains magnetically ordered with gapless excitations
as was the case for the chain. For the chain a gapless nematic phase was identified~\cite{Wang2020b} on either side of the \AK\ phase. For the ladder a similar gapless phase, SN, occurs but
this time only on one side of the \AK\ phase.
\AG\ remains disordered like the chain. There is therefore a close connection between the phases identified in the KG chain but we expect the KG ladder to be much closer to the 
two-dimensional honeycomb lattice and to represent most of the phases occurring in that limit although we in some cases expect phases to become gapless in the two-dimensional limit.
For instance, in the KG ladder \AK\ and \FK\ phases are gapped and we expect these phases to become the AFM and FM
 Kitaev spin liquid, respectively in the 2D limit, which are gapless.


\subsection{Non-zero Field}
When a magnetic field in the [111] direction is introduced, the full $h[111]$, $\phi$ phase diagram is revealed
as shown in Fig.~\ref{fig:FullPD}(a) and (d).  From the rung entanglement, $S_\mathrm{rung}$, shown in Fig.~\ref{fig:FullPD}(d), it is clear that the introduction
of a field in many cases tend to {\it increase} the entanglement. New highly entangled phases appear until the fully field polarized state is attained where
all spins are polarized along $h[111]$.
We denote this polarized state by PS. Trivially, it is a product state with zero entanglement. 
        We start by discussing the fate of the RS$_{U_6}$ phase. This phase is adiabatically connected to the PS phase and no phase transition is observed at any non-zero
field strength. This is contrary to the FM$_{U_6}$ phase state which cannot be adiabatically connected to the PS phase. This follows from the fact that the phase is not an
ordinary FM state but only related to one through the local unitary rotation $U_6$. An alignment of the spins along the [111] direction is therefore energetically costly.
As can be seen in Fig.~\ref{fig:FullPD}(a) and (d) a line
of phase transitions around field strengths of $h[111]=0.7-1.0$ occur, signalling the transition to the PS phase. 
A new phase at finite field above $\phi=\pi/2$ is clearly visible. This is a gapless, magnetically ordered phase with the spins arranged in a zig-zag manner, in opposite directions on the two legs of the ladder, and we refer to this phase
as the ZZ phase. The fate of the phases occurring between
\tcAGtoSNP\ and \tcQPtoFM\ is not known  and left for future study.
The \FK\ and \AG\ phases survive in the presence of a magnetic field and survive up to large field strengths.
The nature of these two phases in
the presence of $h[111]$ has previously been discussed~\cite{Gordon2019}.
We therefore leave that part of the phase diagram aside an instead concentrate on the part of the phase diagram close to the AFM Kitaev point, $\phi=0$.

In the vicinity of $\phi=0$ it is clear from Fig.~\ref{fig:FullPD}(a) and (d) that several new phase are induced by the magnetic field, in many cases
with significantly increased entanglement. The nematic phase, SN,
identified in zero field,  survive in the presence of a non-zero field and is clearly visible in Fig.~\ref{fig:FullPD}(a) and (d). However, the high through-put
iDMRG calculations used in Fig.~\ref{fig:FullPD} are in certain regions having trouble achieving convergence as can be seen by the noise in the figure and a precise
determination of the phase diagram in the vicinity of $\phi=0$ is difficult from the data presented in Fig.~\ref{fig:FullPD}(a) and (d). We therefore
focus on a high resolution study of that part of the phase diagram combined with finite-size DMRG calculations for the regions where it is not possible to achieve
good convergence using iDMRG.


\begin{figure*}
  \centering
  \includegraphics[width=\textwidth]{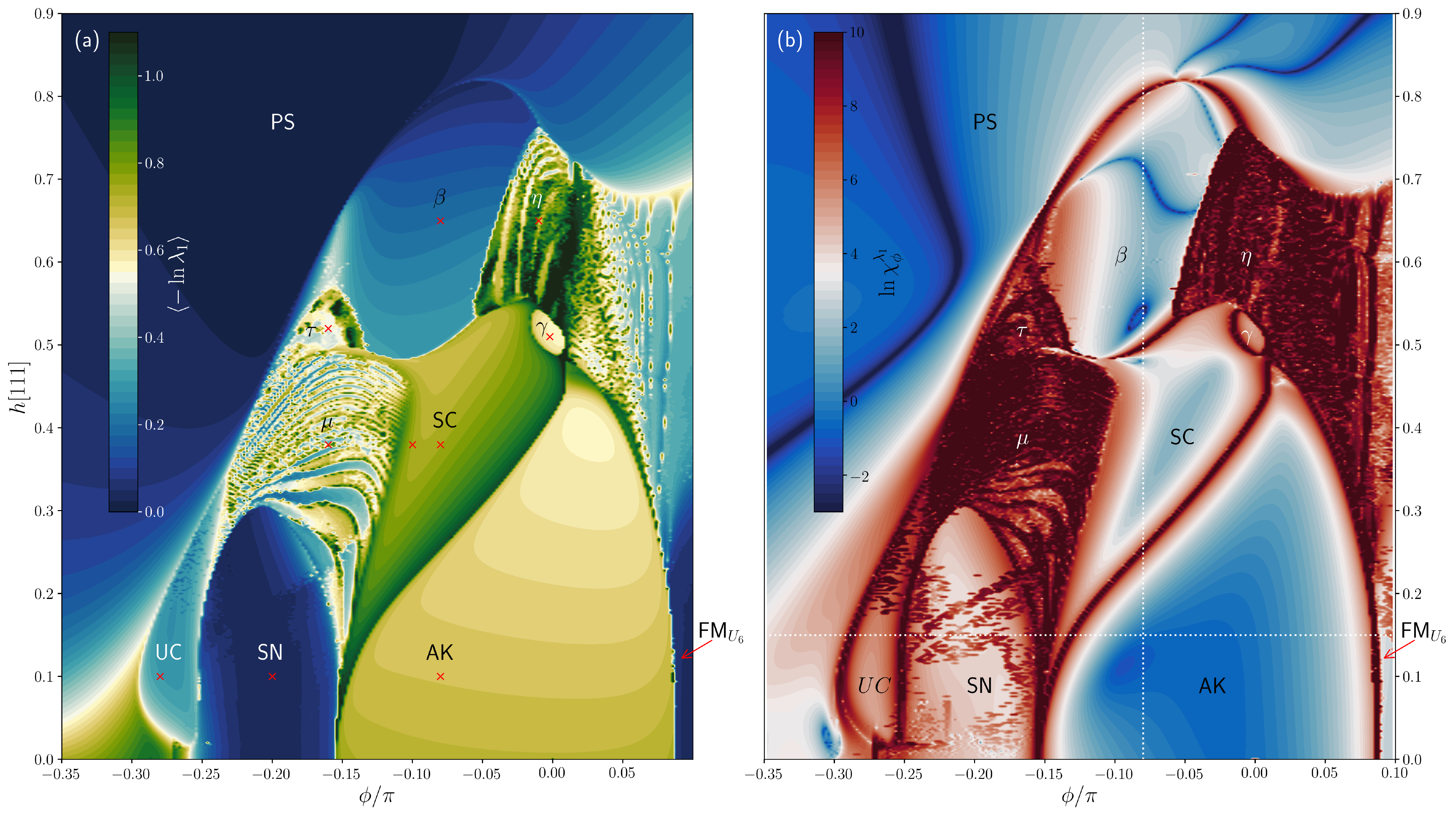}
  \caption{Phase diagram for the AFM Kitaev region of the two-leg ladder KG ladder model under the magnetic field along [111]-axis.
  {\bf (a)} Lowest edge of the entanglement spectrum, $-\ln\lambda_1$ of the reduced density matrix, $\rho_{L/2-1}$, versus $\phi$ in zero field. Dark blue corresponds to $\lambda_1=1 (-\ln\lambda_1=0)$ a phase of low entanglement and
  close to a product state. Dark green corresponds to a small $\lambda_1$ and higher entanglement. The {\color{red}$\times$}'s corresponds to points studied in detail in Fig.~\ref{fig:AFK1Incomm},~\ref{fig:AFK2Incomm}.
  {\bf (b)} The entanglement spectrum susceptibility $\chi_\phi^{\lambda_1}$ on a logarithmic scale.
  Deep blue coloring corresponds to a stable $\lambda_1$ a well defined phase, dark red coloring signals rapid change in $\lambda_1$ and a likely associated phase transition.
  The dashed lines are studied in Fig.~\ref{fig:AFHSweep08},~\ref{fig:AFPSweep15}. Identical phase diagrams obtained from $\chi_h^e$ and $S_\mathrm{rung}$ can be found Appendix~\ref{app:chiesrung}.
  All results in {\bf (a)} and {\bf (b)} are from high throughput iDMRG calculations with a unit cell of 24 sites, with $\Delta\phi=0.002\pi,\ \Delta h[111]=0.002$.
  }
  \label{fig:FineAFK}
\end{figure*}
\begin{figure}[t]
  \centering
        \includegraphics[width=0.5\textwidth]{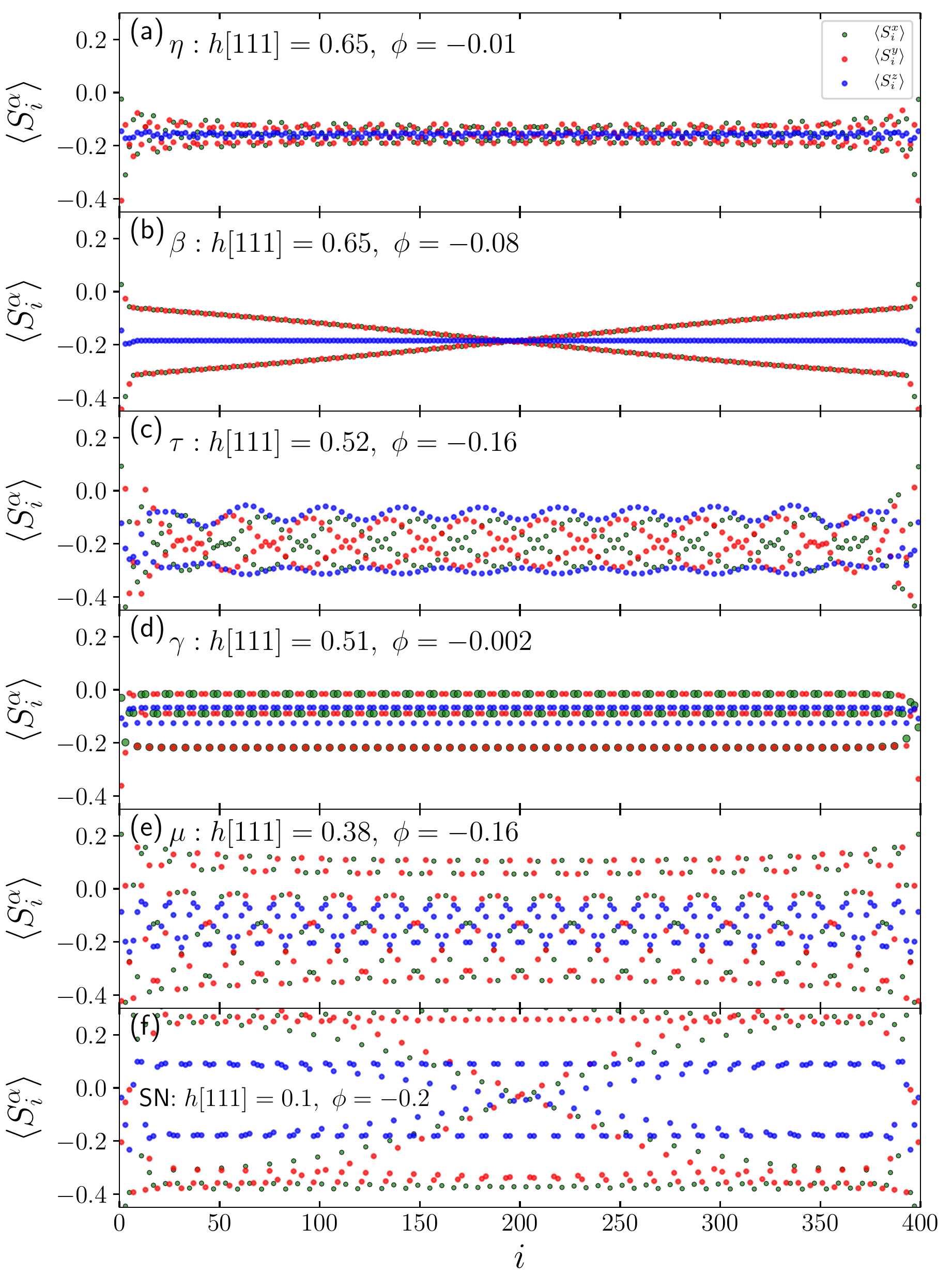}
  \caption{On-site magnetization $\langle S_i^\alpha\rangle$, $\alpha=x$ (green), $y$ (red) and $z$ (blue) along the ladder at six different points in the phase diagram indicated in Fig.~\ref{fig:FullPD} representing
  the (a) $\eta$, (b) $\beta$, (c) $\tau$, (d) $\gamma$, (e) $\mu$ and (f) \SN\ phases.
  Results are from finite-size DMRG calculations for total system size $N=400$ with open boundary conditions. For clarity, results are shown only for {\it one leg} of the ladder.
  }
  \label{fig:AFK1Incomm}
\end{figure}

\begin{figure}[t]
  \centering
        \includegraphics[width=0.5\textwidth]{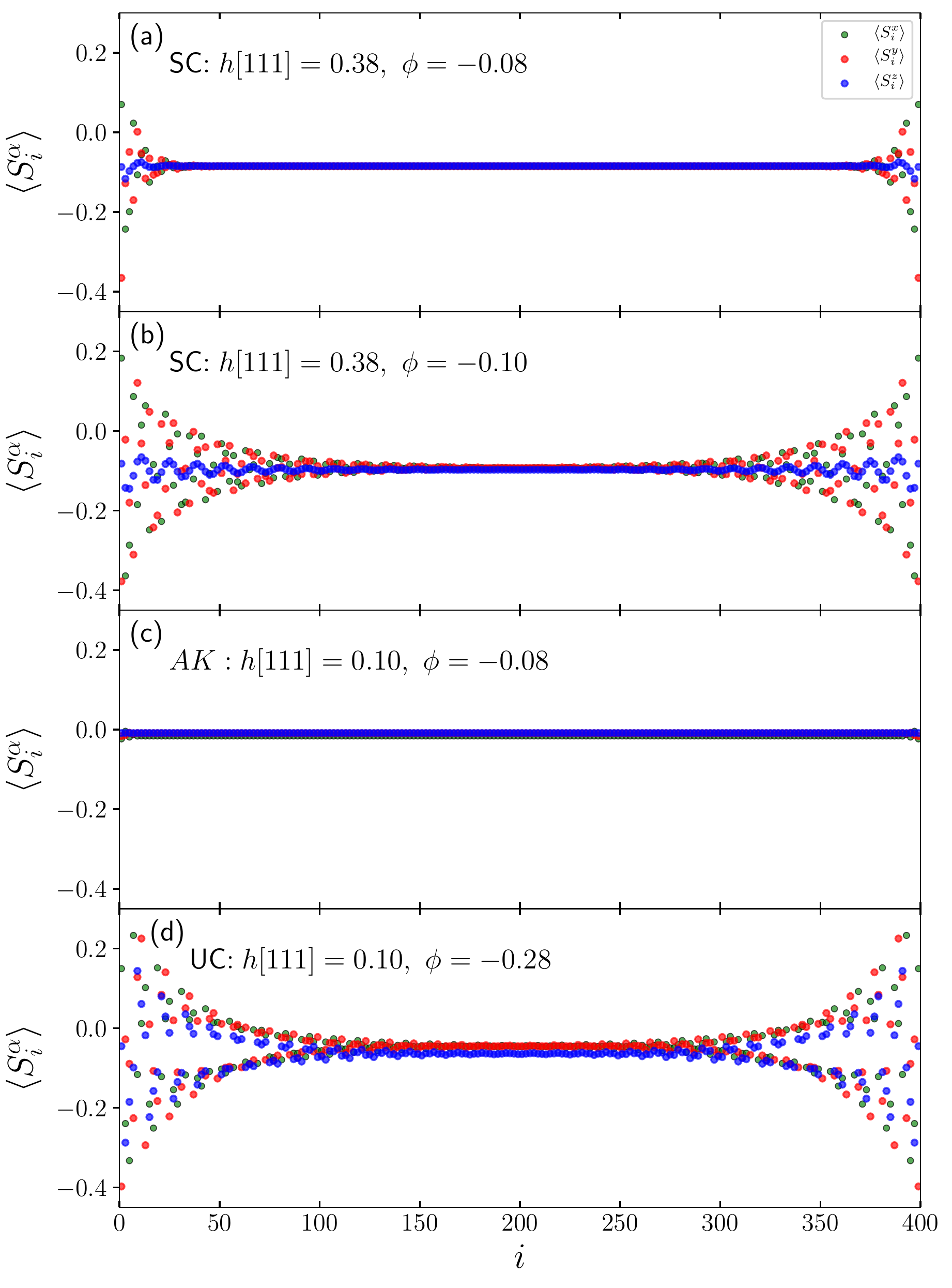}
  \caption{On-site magnetization $\langle S_i^\alpha\rangle$, $\alpha=x$ (green), $y$ (red) and $z$ (blue) along the ladder at four different points in the phase diagram indicated in Fig.~\ref{fig:FullPD} representing
  the (a) SC-phase at $h[111]=0.38$, $\phi=-0.08\pi$, (b) SC-phase at $h[111]=0.38$, $\phi=-0.10\pi$, (c) \AK-phase and (d) UC phase. 
  Results are from finite-size DMRG calculations for total system size $N=400$ with open boundary conditions. For clarity, results are shown only for {\it one leg} of the ladder.
  }
  \label{fig:AFK2Incomm}
\end{figure}

\section{The Antiferromagnetic Kitaev Region, $|\Gamma|\ll K$}\label{sec:afm}

\begin{figure}
  \centering
        \includegraphics[width=0.5\textwidth]{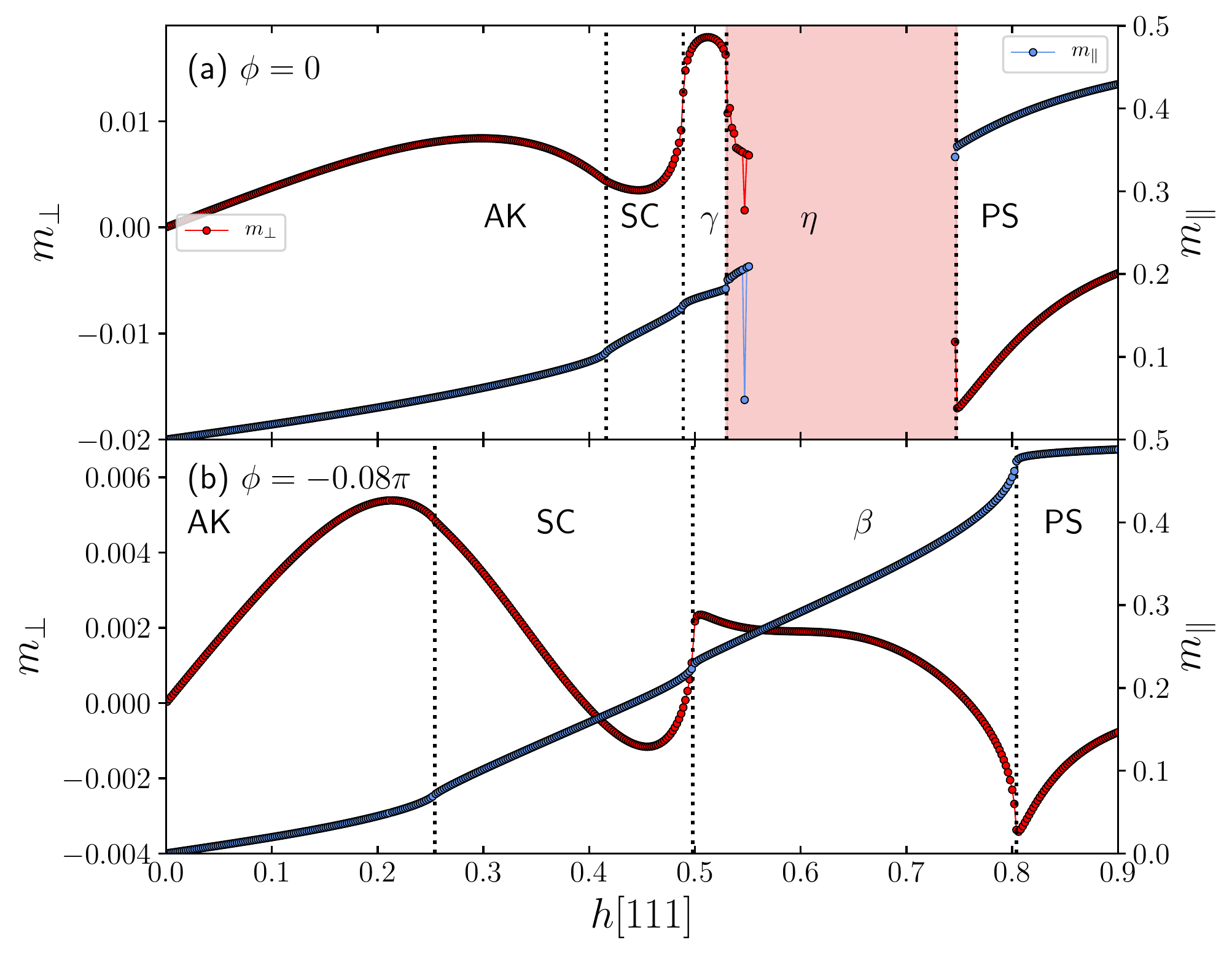}
  \caption{Magnetization versus field, $h[111]$. {\bf (a)} $m_\perp$ (red circles) and $m_\parallel$ (blue circles) versus $h[111]$ at the AFM Kitaev point $\phi=0$.
  {\bf (b)}  $m_\parallel$ (red circles) and $m_\perp$ (blue circles) versus $h[111]$ (blue circles) near the AFM Kitaev point $\phi=-0.08\pi$.
  Results in {\bf (a), (b)} are from high precision iDMRG calculations with a unit cell of 60 performed at $\phi=-0.08\pi$.
  The dotted vertical lines denote the transitions at critical fields determined from divergences in $\chi_h^e$ and $\chi_h^{\lambda_1}$.
  Note that, $m_\parallel$ is roughly 2 orders of magnitude larger than $m_\perp$.
  }
  \label{fig:Magnetization}
\end{figure}

%

The most exotic  part of the phase diagram of the KG model under the field is the AFM Kitaev region around $\phi=0$ 
where $|\Gamma|\ll K$. 
In zero field we have identified the \AK\ spin liquid phase and the nematic phase \SN. However, it is clear that there is a very fine
balance among the different couplings and the presence of a magnetic field will substantially increase this competition. 
We therefore investigate this part of the phase diagram with extreme care. Our results are presented below.

In order to get a detailed picture of the phase diagram we have performed high through-put iDMRG calculations in the region $\phi\in [-0.35,0.10]$ and $h[111]\in[0.00,1.75]$ on a grid with 
$\Delta\phi=0.002\pi,\ \Delta h[111]=0.002$ using 24 site unit cells. By necessity, we have to use a relatively small unit cell and a maximal bond dimension of 500.
Our results are shown in Fig.~\ref{fig:FineAFK}. In panel {\bf (a)} is shown the lower edge of the entanglement spectrum, $-\ln\lambda_1$, determined from the reduced density matrix
that cuts a rung. Phases with negligible entanglement ($\lambda_1\sim 1)$ will show as dark blue where as highly entangled phases ($\lambda_1\ll 1)$ are colored yellow to dark green. The \AK\
phase is clearly visible as the bright yellow phase centered around $\phi=0$, extending to non-zero fields. On the other hand, the \SN\ phase which is close to a product state is clearly
visible as dark blue. In addition to these two previously discussed phases there is a proliferation of phases occurring at higher fields. In order to more clearly identify phase transitions we show
$\chi_\phi^{\lambda_1}$ in Fig.~\ref{fig:FineAFK}(b) on a logarithmic scale.
Deep blue coloring in Fig.~\ref{fig:FineAFK}(b) corresponds to a stable $\lambda_1$, a well defined phase, while dark red coloring signals rapid change in $\lambda_1$ and a likely associated phase transition.
Note the logarithmic scale, where the darkest red coloring is more than 10 orders of magnitude larger than the blue colors. A significant advantage analyzing the phase diagram in the way shown in
Fig.~\ref{fig:FineAFK}(a) and (b), is that the entanglement spectrum, and therefore also $\lambda_1$, {\it has} to change at a quantum phase transition. 

In Fig.~\ref{fig:FineAFK}(b) many clear phase transitions are visible as dark red lines. However, there are also some extended regions with dark red coloring or noise appearing. These regions marked,
$\eta, \tau$ and $\mu$ in Fig.~\ref{fig:FineAFK}(a), (b), are regions where the small unit cell idmrg calculations have difficulty reaching good convergence. Likely this is due to incommensurability
effects and a further investigation using finite size DMRG or ED is warranted. From the data in Fig.~\ref{fig:FineAFK}(a), (b) we identify six well defined phases, the previously discussed \AK\ and \SN\
phases and 4 new phases that we name SC, $\beta$, $\gamma$ and UC. Here SC and UC refer staggered chirality (SC) and uniform chirality (UC) because these phases exhibit
a staggered and uniform pattern of chirality without any magnetic ordering, respectively. 
 As we discuss in more detail below, it appears that the $\mu$ region is distinct from the SC phase and it seems quite
plausible that the $\eta, \tau$ and $\mu$ regions are in fact well-defined phases and we therefore discuss them as such below.  However, at present we cannot exclude the possibility that, for instance,
what appears as a phase transition between the \SN\ phase and the $\mu$ phase is instead a `disorder' line marking the onset of incommensurate short range correlations thereby hindering the convergence of the iDMRG calculations. 
It is therefore possible that the $\eta, \tau$ and $\mu$ regions are not distinct phases but simply parts of the adjoining phases where short-range correlations are different.
We shall return to this point below.
Surprisingly, it is clear that some of the phases occurring at finite field have significantly increased entanglement, most notably the SC phase that together with its accompanying phases, $\gamma$ and $\mu$
outline a heart shaped region of extraordinary high entanglement.

Complementary views of the same phase diagram as obtained from $S_\mathrm{rung}$ and $\chi_h^e$ can be found in Appendix~\ref{app:chiesrung}.

\subsection{Overview of phases, $\langle S^\alpha_i\rangle$}
Several of the phases visible in Fig.~\ref{fig:FineAFK}(a) and (b) are magnetically ordered phases. We have therefore performed high precision finite size
DMRG calculations with $N=400$ and OBC at the points marked with a red $\times$ in Fig..~\ref{fig:FineAFK}(a). Results for the local magnetization $\langle S^\alpha_i\rangle$
are shown in Fig.~\ref{fig:AFK1Incomm} and \ref{fig:AFK2Incomm}, here the green circles are $S^x_i$, the red circles $S^y_i$ and the blue circles $S^z_i$ along
a {\it single} leg of the ladder.
In some cases we find magnetic ordering with a surprisingly large unit cell, in other cases indications of incommensurate ordering.
Note that a simple polarization of the spins along the field direction $h[111]$ would have all $S^\alpha_i$ equal.
Before a more detailed discussion of some of the phases we summarize some of the main findings in Fig.~\ref{fig:AFK1Incomm} and \ref{fig:AFK2Incomm}.
\subsubsection{Magnetically ordered phases}
We start with the six phases that show clear signs of ordering.
\begin{itemize}
  \item $\eta$: This is a high field phase (region). As shown in Fig.~\ref{fig:AFK1Incomm}(a) the local magnetization appear incommensurate. The ground-state is degenerate
     and the phase is likely gapless.
  \item $\beta$: As $\Gamma$ is made slightly more negative the system transitions from the $\eta$ phase to the $\beta$ phase. The local ground-state magnetization along the
    leg of the ladder, shown in Fig.~\ref{fig:AFK1Incomm}(b),  displays a characteristic M\"obius form, showing a single twist in $S^x_i$ and $S^y_i$ from one end of the open chain to the other while $S^z_i$ remains
    constant. The ground-state is degenerate and the phase is possibly gapless. 
  \item $\tau$: This phase (region) is adjacent to the $\beta$ phase but the beautiful intricate ordering along the leg shown in Fig.~\ref{fig:AFK1Incomm}(c) is clearly distinct from that observed in the $\beta$ phase with a large
    variation in $S^z_i$ along the leg. The unit cell for the ordering appears at this value of $\Gamma$ to be approximately 20 lattice spacings along the ladder leg. Again we find a degenerate ground-state and a possible gapless phase.
    At neighboring values of $\Gamma$ we find similar large unit cell ordering.
  \item $\gamma$: The $\gamma$ phase is nested in between the SC and $\beta$ phases occupying a small oval region around $\phi=0$, $h[111]=0.5$. The $\gamma$ phase again show an exquisite ordering along the ladder leg as shown in
    Fig.~\ref{fig:AFK1Incomm}(d), in this case
    with a smaller unit cell of 5 lattice spacings. There is no variation in the ordering throughout the $\gamma$ phase. The ground-state is degenerate and the phase is possibly gapless.
    However, the bipartite entanglement entropy $S(x)$ is close to constant throughout most of the ladder which would suggest a gapped phase (see Appendix~\ref{app:sx}).
  \item $\mu$: This phase (region) is nested above the \SN\ phase. The magnetic ordering is shown in Fig.~\ref{fig:AFK1Incomm}(e), in this case with a unit cell of 13 lattice spacings along the leg of the ladder. Neighboring values of $\gamma$ show
      variations in the local magnetization pattern and it is not clear to what extent this phase is different from  the $\tau$ phase. For the $\mu$ phase we again find a degenerate ground-state and possibly a gapless phase.
    \item \SN: In zero field the \SN\ was clearly identified. There is no indication of a phase transition as the $h[111]$ is introduced and we therefore assume that the well defined uniform phase visible
      in Fig.~\ref{fig:FineAFK}(a) and (b) is adiabatically connected to the \SN\ phase in zero field. The phase is gapless and the presence of the non-zero magnetic field induces an incommensurate local magnetization as shown
      in Fig.~\ref{fig:AFK1Incomm}(f).
\end{itemize}
As outlined above, the large unit cell ordering occurring in the $\mu$ and $\tau$ phases (regions) varies with $\Gamma$ and $h[111]$. The same effect is observed for the incommensurate ordering in the $\eta$ phase. 
In fact, the stripes occurring in these phases visible in Fig.~\ref{fig:FineAFK}(a) are likely caused 
by the variations in the local ordering best compatible with the 24 site unit cell. The lines could therefore represent lock in transitions.

These phases
are therefore not uniform in a conventional sense but could possibly be a series of phases with shifting sizes of unit cells for the magnetic ordering. We have not been able to resolve this.

\subsubsection{Non magnetic phases}
We now turn to a discussion of the three remaining disordered phases which do not show any conventional local magnetic ordering apart from that induced by the magnetic field.
\begin{itemize}
  \item SC: In Fig.~\ref{fig:AFK2Incomm}(a) and (b) is shown the local magnetization in the SC phase at two different values of $\phi$ both at $h[111]=0.38$. In the middle of the chain
    the local magnetization aligns with the $h[111]$ field and the phase is best described as disordered. Clear excitations at the end of the open chain are visible. As $\phi$ is increased from -0.08 to -0.10
    approaching the $\mu$ phase the size of the chain end excitations visibly grow. The SC phase is highly entangled, significanly more so than the \AK\ phase. The ground-state does not appear degenerate but at $h[111]=0.38$, $\phi=-0.05$ we can limit the gap to the first excited
    state by $\Delta_1<0.019$. The gap is likely smaller in other parts of the phase. 
    The bipartite entanglement entropy $S(x)$ is close to constant throughout most of the ladder which is also consistent with a gapped phase (see Appendix~\ref{app:sx}).
    The phase show signs of scalar chiral ordering as we discuss in more detail below.
    \item \AK: This is the AFM  Kitaev phase previously discussed. 
      As shown in In Fig.~\ref{fig:AFK2Incomm}(b) the local magnetization is aligned with the $h[111]$ field and only faint signs of chain end excitations are visible.
      The phase has a gap and in zero field a two fold degenerate ground-state. It is possible to define
      a string order parameter (SOP)~\cite{Catuneanu2018ladder} as we discuss below.
    \item UC: On the left side of the \SN\ phase a new phase appears in the presence of a $h[111]$ field. As we shall discuss below this phase has scalar chiral ordering.
      The ground-state is degenerate and the phase is possibly gapless. However, the bipartite entanglement entropy $S(x)$ is close to constant throughout most of the ladder which would suggest a gapped phase (see Appendix~\ref{app:sx}).
\end{itemize}

\subsubsection{Total magnetization, $m_\perp$ and $m_\parallel$}
It is useful to analyze the total magnetization of the open chain by separating the components parallel, $m_\parallel$, and perpendicular, $m_\perp$, to the $[111]$ field direction.
We define $s^\alpha=\sum_i s^\alpha_i/N$ and then
\begin{equation}
m_\parallel = (s^x+s^y+s^z)/\sqrt{3}
\end{equation}
It follows that
\begin{equation}
  \vec m_\perp=\vec s-m_\parallel(1,1,1)/\sqrt{3}.
\end{equation}
To facilitate visualization it is most convenient to plot $|m_\perp|$ with a sign that we determine as $\mathrm{sign}(m_\perp\cdot \hat a)$ with
$\hat a=(1,1,-2)/\sqrt{6}$. Surprisingly, $m_\perp$ is completely aligned or anti-aligned with $\hat a$ for the two cases we shall now discuss and the angle
$m_\perp$ forms with $\hat a$ in the $\hat a$, $\hat b$ plane is either 0 or $\pi$. Our results are shown in Fig.~\ref{fig:Magnetization}(a),(b) for $\phi=0$ and $\phi=-0.08\pi$, respectively.
The results are from high precision iDMRG calculations with a unit cell of 60 and a maximal bond dimension of 1,000. 
An important point to notice in Fig.~\ref{fig:Magnetization} is that $m_\perp$ is 2-3 orders of magnitude smaller than $m_\parallel$. 

Let us first discuss the results shown in Fig.~\ref{fig:Magnetization}(a) obtained from a field sweep from 0 to 0.9 at $\phi=0$. As the $\eta$ phase is entered the iDMRG calculations
fail to converge and that part of the plot is therefore colored lightly red. Starting from zero field $m_\parallel$ (blue circles) is found to approximately linearly increase with $h[111]$ until
the SC phase is reached where a kink in $m_\parallel$ is observed, consistent with a divergence of $\chi_h^e$. Through the SC phase $m_\parallel$ increase more rapidly, this phase is therefore
'softer', consistent with the large chain end excitations shown in Fig.~\ref{fig:AFK2Incomm}(a),(b). A second kink in $m_\parallel$ is observed as the $\gamma$ phase is entered but the increase in $m_\parallel$ through
out the $\gamma$ phase is less pronounced. As the $\eta$ phase is entered and excited kinks in $m_\parallel$ are again observed and in the PS phase the $m_\parallel$ tend toward the fully polarized value of 1/2. On the other
hand $m_\perp$ (red circles) is non monotonic throughout the \AK\ phase, approaches small values in the SC phase before jumping to larger values in the $\gamma$ phase. In the polarized state (PS) phase $m_\perp$ changes sign
and approach zero from below.

Our results for a similar field sweep at $\phi=-0.08\pi$ is shown in Fig.~\ref{fig:Magnetization}(b). The position of this field sweep is indicated as the vertical dotted line in Fig.~\ref{fig:FineAFK}(b). Again, $m_\parallel$ is seen to increase more rapidly through the SC and $\beta$ phases
as compared to the \AK\ phase. The phase transitions between the \AK\, SC and $\beta$ phases are clearly visible as kinks in $m_\parallel$ and as the PS phase is approached $m_\parallel$ approach
1/2 in a characteristic cusp that was not observed between the $\eta$ and PS phase. The nature of the $\beta$-PS transition and $\eta$-PS transition therefore clearly appear different. 
In this case $m_\perp$ is featureless at the \AK\ to SC transition but the SC$-\beta$ and $\beta-$PS transitions
are clearly visible. $m_\perp$ changes sign in the SC and $\beta$ phases. However, $m_\perp$ is in this case 3 orders of magnitude smaller than $m_\parallel$.

For both field sweeps at $\phi=0$ and $\phi=-0.08\pi$ we emphasize that $m_\perp$ is either aligned or anti-aligned with $\hat a$.

\begin{figure}
  \centering
        \includegraphics[width=0.5\textwidth]{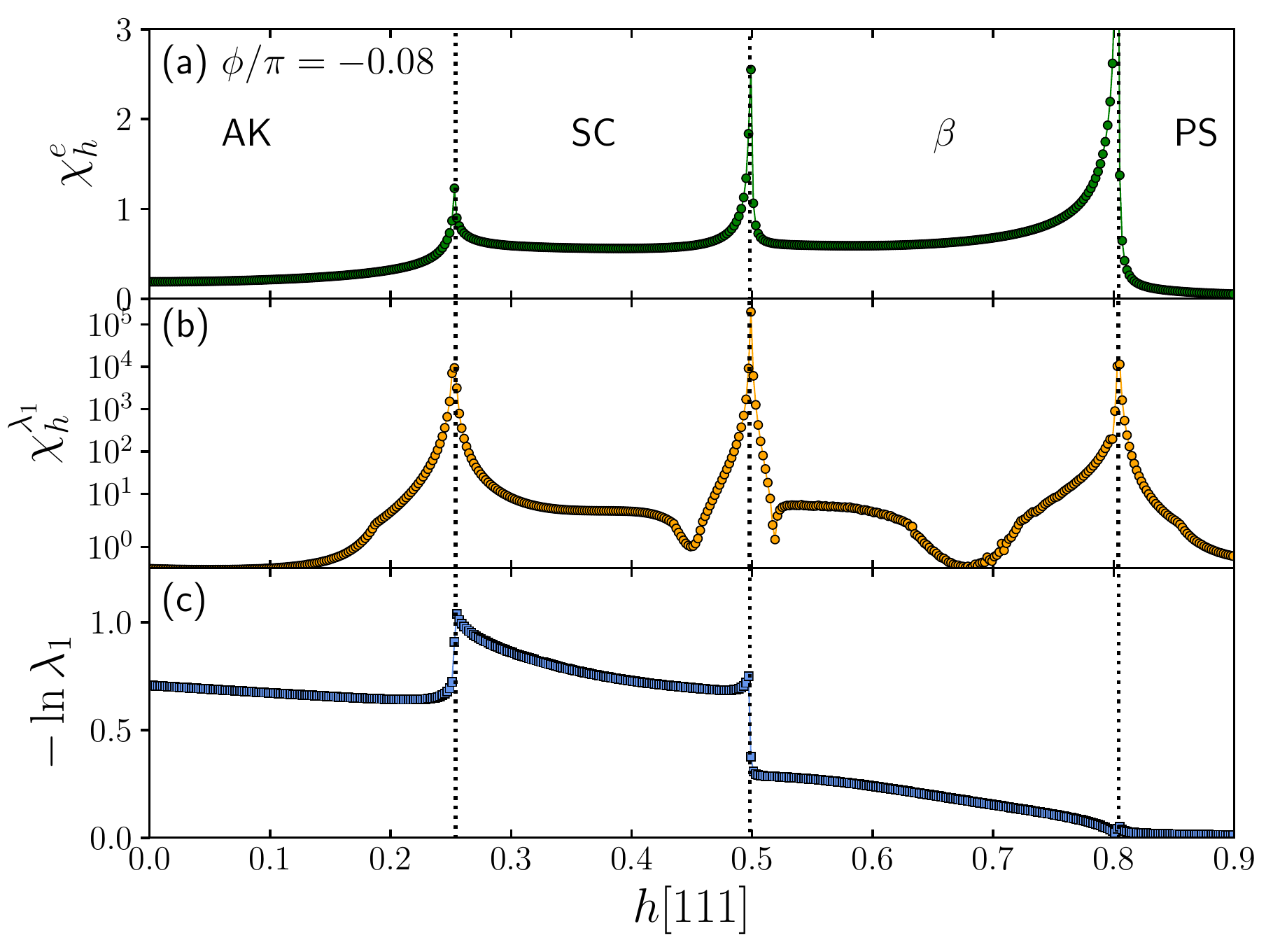}
  \caption{Field sweep of the phase diagram in the AFM Kitaev region at $\phi=-0.08\pi$. 
  {\bf (a)} $\chi_h^e$ versus $h[111]$. {\bf (b)} $\chi_h^{\lambda_1}$ versus $h[111]$, note the logarithmic scale. {\bf (c)} Lowest edge of entanglement spectrum, $-\ln\lambda_1$ from $\rho_{L/2-1}$ versus $h[111]$.
  All results in {\bf (a)}, {\bf (b)} and {\bf (c)} are from high precision iDMRG calculations with a unit cell of 60. The dashed vertical lines correspond to transitions at field values of $h[111]$ of $0.254$(AK-SC), $0.498$ (SC-$\beta$) and $0.804$ ($\beta$-PS).
  }
  \label{fig:AFHSweep08}
\end{figure}

\begin{figure}
  \centering
        \includegraphics[width=0.5\textwidth]{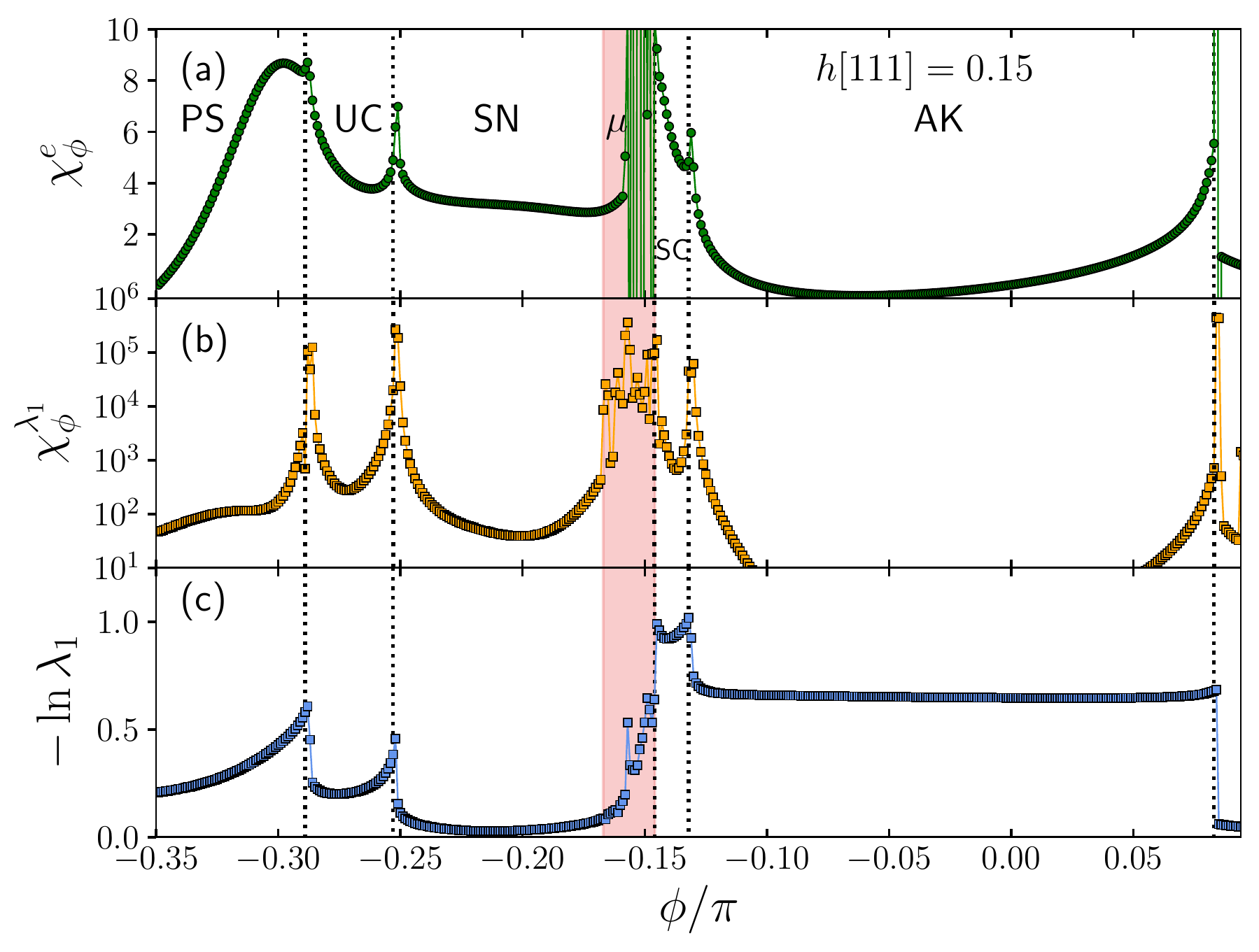}
  \caption{Angle sweep of the phase diagram in the AFM Kitaev region at $h=0.15$. 
  {\bf (a)} $\chi_\phi^e$ versus $\phi/\pi$, note the almost imperceptible divergence in $\chi_\phi^e$ at $\phi=-0.289\pi$. {\bf (b)} $\chi_\phi^{\lambda_1}$ versus $\phi/\pi$, note the logarithmic scale. {\bf (c)} Lowest edge of entanglement spectrum, $-\ln\lambda_1$ from $\rho_{L/2-1}$ versus $\phi/\pi$.
  All results in {\bf (a)}, {\bf (b)} and {\bf (c)} are from high precision iDMRG calculations with a unit cell of 60. The dashed vertical lines correspond to transitions at angles $\phi/\pi$ of $-0.289$(PS-UC), $-0.253$ (UC-\SN), $-0.146$ ($\mu$-SC), $-0.132$ (SC-\AK) and $0.083$ (\AK-FM$_{U_6}$).
  The light red coloring denotes a region of $\phi/\pi$ where convergence of the iDMRG is problematic.
  }
  \label{fig:AFPSweep15}
\end{figure}

\subsubsection{Field and angle sweeps}
To further investigate the sharpness of the phase transitions occurring in Fig.~\ref{fig:FineAFK} we have performed high precision sweeps 
at constant $\phi=-0.08\pi$ and constant $h[111]=0.15$. The positions of these sweeps are shown as the dotted lines in Fig.~\ref{fig:FineAFK}(b).
Our results are shown in Fig.~\ref{fig:AFHSweep08} for constant $\phi=-0.08\pi$ and in Fig.~\ref{fig:AFPSweep15} for constant $h[111]=0.15$. In both
case from high precision iDMRG calculations with a unit cell of 60 and a maximal bond dimension of 1,000.

Results for a field sweep from $h[111]=0$ to $0.9$ are shown in Fig.~\ref{fig:AFHSweep08} for a constant $\phi=-0.08\pi$. In panels (a), (b) and (c)  we show
results for $\chi_h^e$, $\chi_h^{\lambda_1}$ and $-\ln\lambda_1$, respectively. The 3 transitions, \AK-SC ($h[111]$=0.254), SC-$\beta$ ($h[111]=0.498$) and $\beta$-PS ($h[111]=0.804$) are very well
defined and are in complete agreement between the three different measures. While the $\beta$-PS is not all that visible in $-\ln\lambda_1$ it is very clear in $\chi_h^{\lambda_1}$. Five orders of magnitude
variation is observed in $\chi_h^{\lambda_1}$.

Results for a $\phi$ sweep from $\phi=-0.35\pi$ to $\phi=0.1\pi$ at constant $h[111]=0.15$ are shown in Fig.~\ref{fig:AFPSweep15}. In panels (a), (b) and (c)  we show
results for $\chi_h^e$, $\chi_h^{\lambda_1}$ and $-\ln\lambda_1$, respectively. 
Five transitions are clearly visible. While the PS-UC transition at $\phi=-0.289\pi$ is easy to miss in $\chi_h^e$ it is very well defined in $\chi_h^{\lambda_1}$ and 
$-\ln\lambda_1$. However, there is a precursor peak in $\chi_h^e$ not associate with a transition (see also Appendix~\ref{app:chiesrung}). 
The transition SC-\AK\ is clearly defined at $\phi=-0.132\pi$ as the $\mu$ phase is approached from the SC side it is also clear that
$\chi_h^e$, $\chi_h^{\lambda_1}$ diverge at 
the $\mu$-SC transition at $\phi=-0.146\pi$. This transition is therefore well defined. On the other hand, the iDMRG fail to achieve good convergence in the light red colored region so the
transition from \SN\ to $\mu$ is not clear. As discussed above it is possible that the $\mu$ phase only marks the onset of short range incommensurate correlations in the \SN\ phase and it is
not a distinct phase. It is also possible that improved convergence of the iDMRG would show a divergence in 
$\chi_h^e$ and $\chi_h^{\lambda_1}$ inside the lightly red colored region.

\begin{figure}[t]
  \centering
        \includegraphics[width=0.5\textwidth]{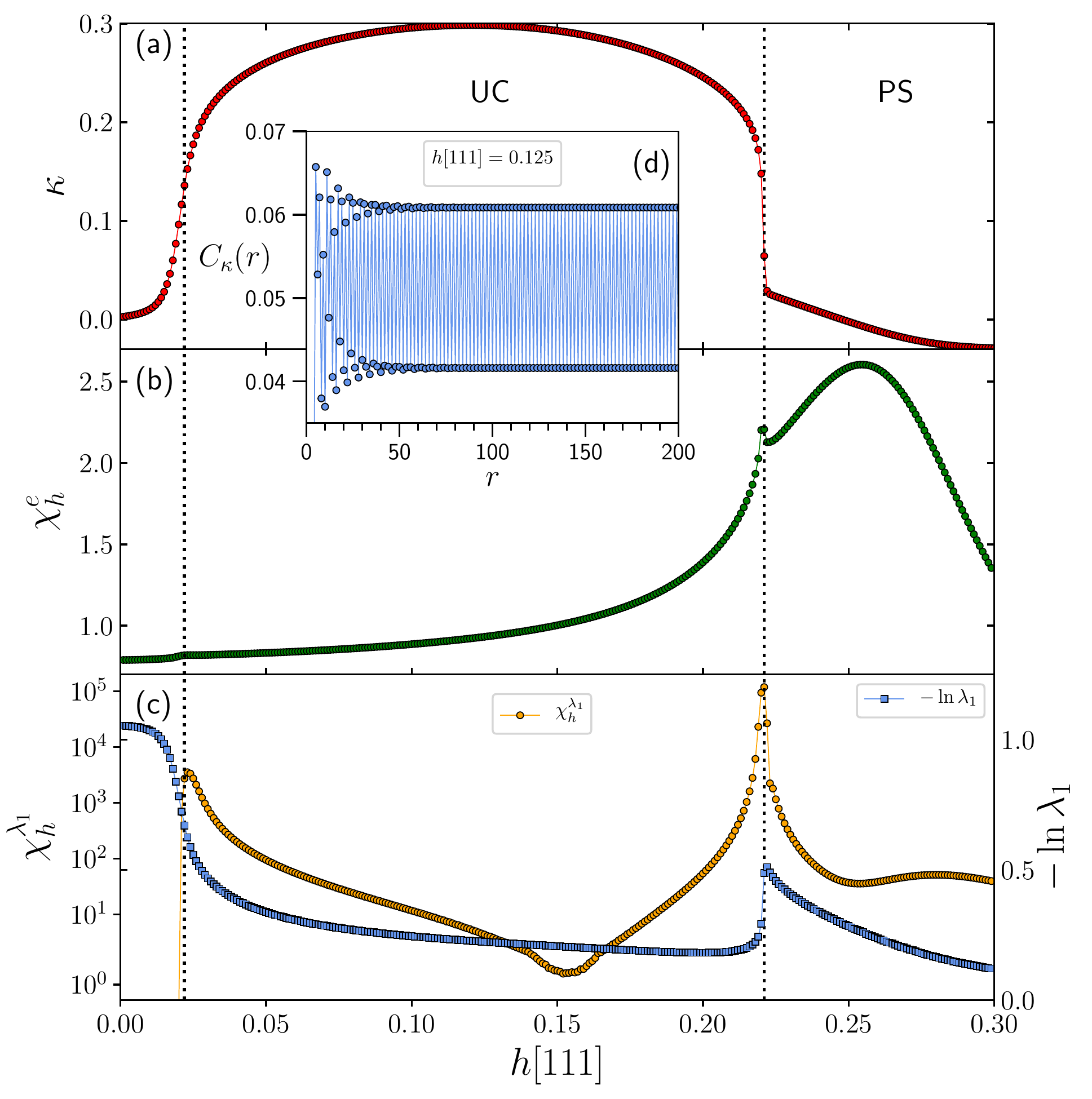}
  \caption{Ordering in the UC-phase. {\bf (a)} The scalar chirality $\kappa$ versus $h[111]$.
  {\bf (b)} $\chi_h^e$ versus $h[111]$. {\bf (c)} $\chi_h^{\lambda_1}$ (orange circles) and $-\ln\lambda_1$ (blue circles) versus $h[111]$.
  {\bf (d)} The chiral correlation function, $C(r)$ along the ladder at $h[111]=0.125$.
  All results in {\bf (a - d)} are from high precision iDMRG calculations with a unit cell of 60 performed at $\phi=-0.27\pi$.
  The dotted vertical lines denote the transitions at critical fields $h_{1}^c=0.022$ and $h^c_2=0.221$.
  Note that, the divergence in $\chi_h^e$ at $h_{1}^c$ is imperceptible and very small at $h^c_2$ where it occurs away from the maximum.
  However, $\chi_\phi^{\lambda_1}$ show very well defined peaks at both $h_{1}^c$ and $h^c_2$.
  }
  \label{fig:Chiraldelta}
\end{figure}

\begin{figure}[t]
  \centering
        \includegraphics[width=0.5\textwidth]{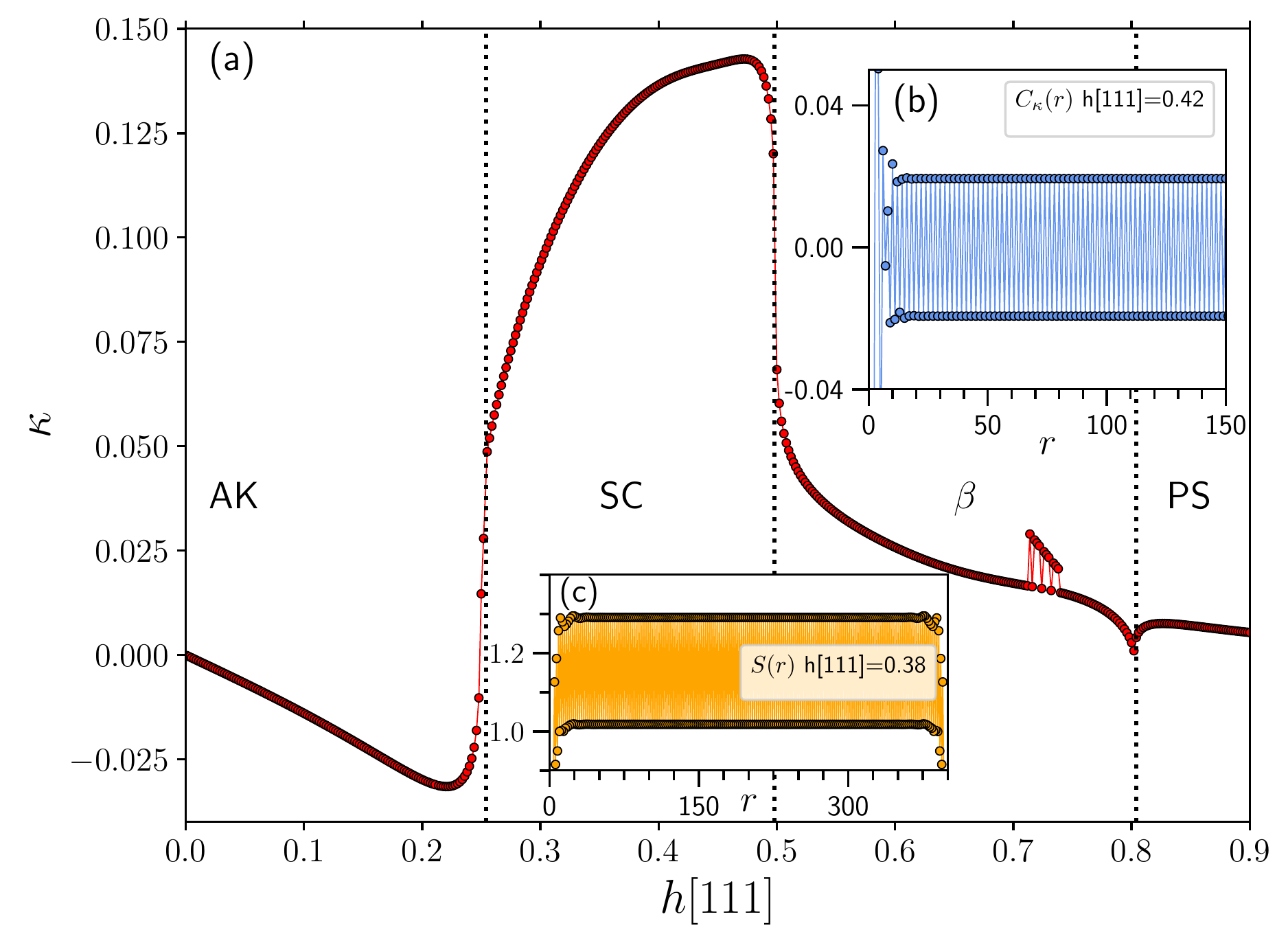}
  \caption{Ordering along $\phi=-0.08\pi$ (AK-, SC- and $\beta$-phase). {\bf (a)} The scalar chirality $\kappa$ versus $h[111]$.
  {\bf (b)} The chiral correlation function, $C(r)$ along the ladder at $h[111]=0.42$.
  {\bf (c)} Bipartite entanglement entropy, $S(r)$ versus $r$ at $h[111]=0.38$.
  Results in {\bf (a), (b)} are from high precision iDMRG calculations with a unit cell of 60 performed at $\phi=-0.08\pi$ while
  {\bf (c)} is from finite-size DMRG with OBC at $\phi=-0.08\pi$, $h[111]=0.38$.
  The dotted vertical lines denote the transitions at critical fields $h_{1}^c=0.254$, $h^c_2=0.498$ and $h^c_3=0.804$.
  }
  \label{fig:Chiralalpha}
\end{figure}

\subsection{Scalar Chirality in the UC and SC phases}
The presence of a non-zero $\Gamma$ term or the magnetic field raises the possibility of chiral ordering. 
The chirality without magnetic ordering is rare, unless there are three or four-spin interactions.
In 1D system, it was shown that a four-spin interaction produces a long-range scalar chirality.\cite{Lauchli2003}
To check the presence of chiral ordering,  we label
the $i$'th spin on two-legs of the ladder as ${\bf S}_{i,1}$ and ${\bf S}_{i,2}$ where $1$ and $2$ refer
to the bottom-leg and top-leg, respectively.
We then define the scalar
chiral order parameter with ${\bf S}={\bf\sigma}/2$ as follows.
\begin{equation}
  \kappa = \langle {\bf \sigma}_{i,1}\cdot \left({\bf \sigma}_{i,2}\times {\bf \sigma}_{i+1,1}\right)\rangle.
\end{equation}
This clockwise definition is kept for all triangles made of three spins,
 for example $\kappa = \langle {\bf \sigma}_{i,2}\cdot \left({\bf \sigma}_{i+1,2}\times {\bf \sigma}_{i+1,1}\right)\rangle$ for upper triangles.
If the $\kappa$ is positive (negative), we assign blue (red) arrows $i\to j\to k$ for $\kappa=\langle {\bf \sigma}_{i}\cdot \left({\bf \sigma}_{j}\times {\bf \sigma}_{k}\right)\rangle$ and all even permutations of $i,j,k$, which leads to the clockwise (anti-clockwise) circulation. 

It is also of interest to define the scalar chiral correlation function,
\begin{equation}
  C_\kappa(r) = \langle \kappa_i \kappa_{i+r}\rangle,
\end{equation}
such that $C_\kappa(r)\to\kappa^2$. The scalar chiral order parameter breaks spatial symmetries and time reversal symmetry,
but not SU(2).

In Fig.~\ref{fig:Chiraldelta}(a) we show $\kappa$ along a field sweep at constant $\phi=-0.27\pi$ through the UC phase. 
Results are from high precision iDMRG with a unit cell of 60 sites and a maximal bond dimension of 1,000. In this case we 
determine $\kappa$ from $C_\kappa(r)$ shown in the inset Fig.~\ref{fig:Chiraldelta}(d) at $h[111]=0.125$. $C_\kappa(r)$
reaches a constant value relatively quickly and $\kappa$ is clearly
non-zero throughout the phase, reaching significant values at the center of the UC phase. The scalar chirality, $\kappa$
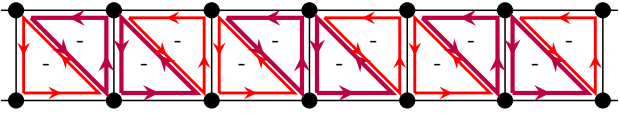
\begin{figure}[h!]
  \centering
  \begin{tikzpicture}[scale=1]
    \chrldelta{6}
   \end{tikzpicture}
  \caption{ Scalar chirality in the UC-phase. At $h = 0.125$ and $\phi=-0.27$,
  $\kappa = -0.204, -0.298, -0.298, -0.204$ for first four triangles respectively, and then repeat. }
  \label{fig:kappadelta}
\end{figure}
is negative throughout the chain and a sketch of the spatial modulation is shown in Fig.~\ref{fig:kappadelta}. 
Note the edge states appearing on the two opposing legs.
Here weaker lines indicate a weaker $\kappa$.
Since the $h[111]$ field favors alignment of the spins, which would result in $\kappa=0$, it
is rather surprising to observe such a well defined scalar chirality at finite fields.
In Fig.~\ref{fig:Chiraldelta}(b), (c)
are shown $\chi_h^e$  and $\chi_h^{\lambda_1}$, respectively. 
While the transitions delimiting the UC phase are almost absent in $\chi_h^e$, and
clearly do not coincide with the broad maximum of $\chi_h^e$ around $h[11]\sim0.25$, they
are very well defined in $\chi_h^{\lambda_1}$ and in both cases they coincide with the results for $\kappa$ in Fig.~\ref{fig:Chiraldelta}(a). 
As far as we can tell the UC phase does not intersect the zero field axis, instead, as can be seen in Fig.~\ref{fig:FineAFK} the
UC and \SN\ phases meet at a triple point close to $\phi=-0.265\pi$. Entanglement close to this triple point is therefore very elevated
which is why $-\ln\lambda_1$ (blue squares in Fig.~\ref{fig:Chiraldelta}(c)) is so high close to zero field.

\begin{figure}[h!]
  \centering
  \begin{tikzpicture}[scale=1]
    \chrlalpha{6}
   \end{tikzpicture}
  \caption{Scalar chirality in the SC-phase. At $h = 0.42$ and $\phi=-0.08$, $\kappa = 0.139, -0.139, -0.139, 0.139$ for first four triangles respectively, and then repeat.}
  \label{fig:kappaalpha}
\end{figure}
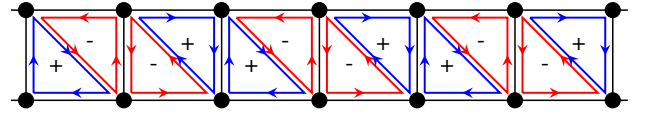

In Fig.~\ref{fig:Chiralalpha} we show iDMRG calculations for $\kappa$ versus $h[111]$ at a fixed $\phi=-0.08$, crossing the \AK, SC and
$\beta$ phases. $\chi_h^e$ and $\chi_h^{\lambda_1}$ along the same line in the phase diagram are shown in Fig.~\ref{fig:AFHSweep08}. As before,
$\kappa$ is obtained from calculations of $C_\kappa(r)$ in the large $r$ limit and the sign of $\kappa$ from local direct estimates of $\kappa$.
In this case there is a spatial +- alternation of $\kappa$ as shown in Fig.~\ref{fig:kappaalpha} but the magnitude of $\kappa$ is the same for each triangle
leading to the zero flux in the system. 
Although there is weak chirality in the \AK\ and $\beta$ phases, $\kappa$ is an order of magnitude larger in the SC phase and jumps
rather abruptly at the critical points indicated by the dotted lines in Fig.~\ref{fig:Chiralalpha}. In the inset panel (b) is shown
$C_\kappa(r)$ versus $r$ at $h[111]=0.42$ which attain a constant value for modest values of $r\sim 20$. Panel (c) in Fig.~\ref{fig:Chiralalpha}
show the bipartite entanglement obtained from a bi-partition of the system at site $r$. The resulting entanglement entropy $S(r)=-\rm{Tr}\rho_r\ln\rho_r$
is shown versus $r$ for a finite ladder with $N=400$ and OBC. Clearly, $S(r)$ is close to constant in the middle of the ladder which would be consistent
with the existence of a non-zero gap in the SC phase. (See also Appendix~\ref{app:sx}).

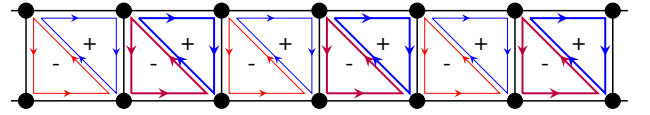
\begin{figure}[h!]
  \centering
  \begin{tikzpicture}[scale=1]
    \chrlAK{6}
   \end{tikzpicture}
  \caption{Scalar chirality in the \AK-phase. At $h=0.2$ and $\phi = -0.08$, $\kappa= -0.0191, 0.0191, -0.0466, 0.0466$ 
  for first four triangles respectively,  and then repeat. }
  \label{fig:kappaAK}
\end{figure}

To make a comparison to the \AK\ phase, we also compute the chirality in the \AK\ phase, and its pattern is shown in Fig.~\ref{fig:kappaAK}. 
$\kappa$ has distinctly different circulations. It has a staggering circulation between upper and lower triangles, but different staggering
from  the SC phase. For both the \AK\ and SC phases, the magnitude of $\kappa$ is the same for all pairs of triangles,
implying that there is zero net-flux in the system. The phase transition from the \AK\ to SC phases is accompanied by
the sharp change of both magnitude and distinct pattern of $\kappa$.

\subsection{String Order and Mapping to KQ Model}
In Ref.~\cite{Catuneanu2018ladder} a {\it non-local} unitary transformation, $V$, was introduced of the following form:
We then define the following unitary (disentangling) operator for an $N$-site chain with OBC:
\begin{equation}
  V = \prod_{\substack{j+1< k \\ j\ \mathrm{odd},\  k\ \mathrm{odd} \\ j=1,\ldots N-3 \\ k=3,\ldots N-1}} U(j,k).
\end{equation}
With the individual $U(j,k)$ given as follows:
\begin{equation}
  U(j,k)  = e^{i\pi(S^y_j+S^y_{j+1})\cdot(S^x_k+S^x_{k+1})}.
\end{equation}
At the AFM Kitaev point, $\phi=0$, $V$ maps the ladder with open boundary conditions to a so called dangling-Z model, $H_{d-Z}$,
that has long-range ordering in $\langle \tilde S^z_i \tilde S^z_{i+r}\rangle$, where $\tilde S$ are the spins in $H_{d-Z}$~\cite{Catuneanu2018ladder}.
If this correlation function is transformed back to the original Kitaev ladder one arrives at a string correlation function of the following 
form:
\begin{eqnarray}
  \langle{\cal O}^z(r)\rangle &=& 4 \langle \tilde S_2^z\tilde S_{2+r}^z\rangle =  (-1)^{\left \lfloor{(r+1)/2}\right \rfloor }\nonumber\\
  &\times&\left\{ \begin{array}{rl}
  \langle\sigma_1^y\sigma_2^x \left( \prod_{k=3}^{r}\sigma_k^z \right)\sigma_{r+1}^x\sigma_{r+2}^y\rangle &\mbox{ $r$ even}\\
    \\
  \langle\sigma_1^y\sigma_2^x \left( \prod_{k=3}^{r+1}\sigma_k^z \right)\sigma_{r+2}^y\sigma_{r+3}^x\rangle &\mbox{ $r$ odd}
       \end{array} \right.
       \label{eq:oz}
\end{eqnarray}
Note that, in Ref.~\cite{Catuneanu2018ladder} some of the indices in Eq.~(\ref{eq:oz}) in the expression for $r$ even were incorrect. With this definition 
we find that the usual plaquette operator~\cite{kitaev2006} $W_p\equiv{\cal O}^z(r=4)$. $W_p$ is often used to characterize the \AK\ phase. See Appendix~\ref{app:wp}.

\begin{figure}[t]
  \centering
        \includegraphics[width=0.5\textwidth]{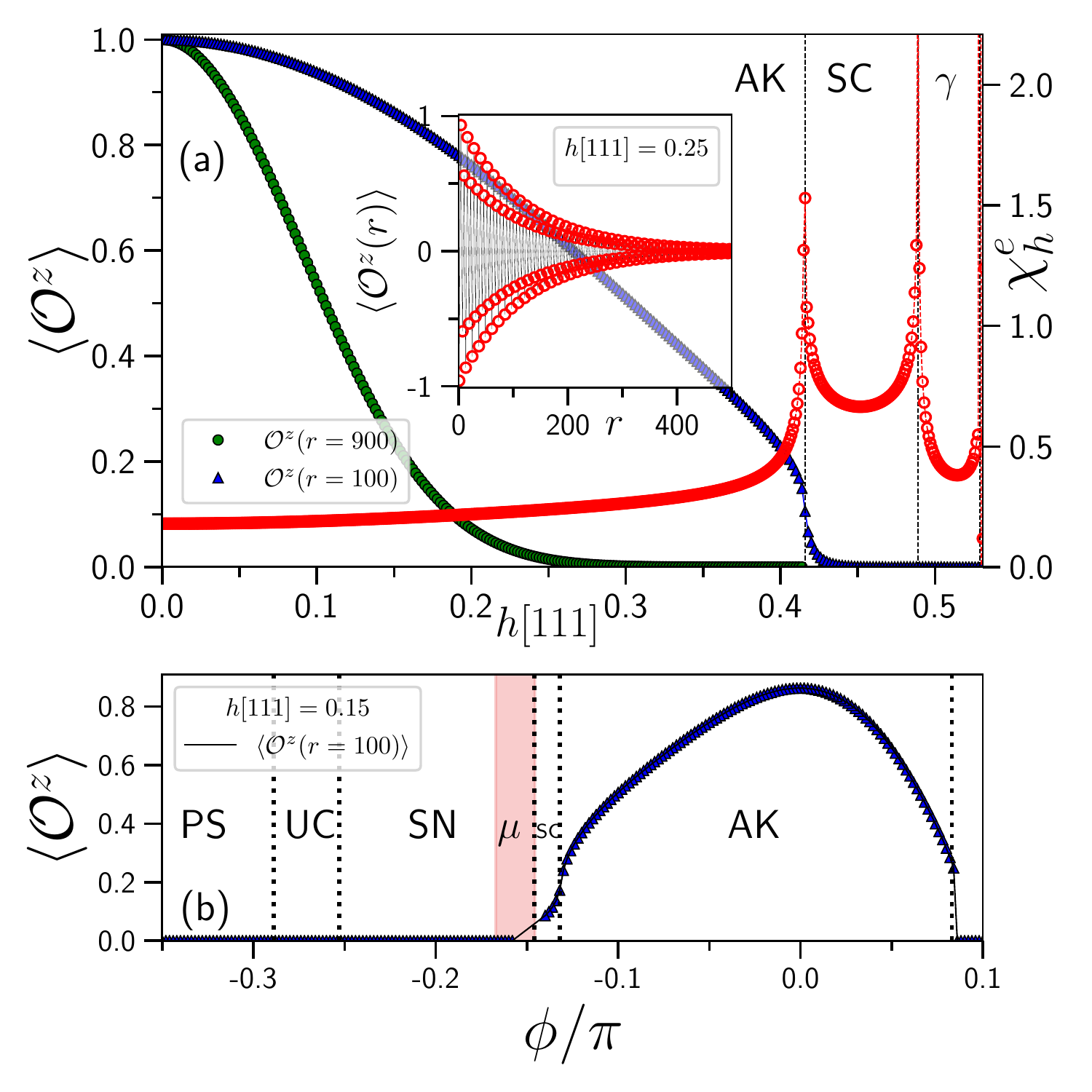}
  \caption{String order $\langle {\cal O}^z\rangle$ versus field in the AFM Kitaev region.
  {\bf (a)} $\langle {\cal O}^z(r=100)\rangle$ (blue triangles) and $\langle {\cal O}^z(r=100)\rangle$ (green circles) versus $h[111]$ at the AFM Kitaev point $\phi=0$ shown
  along with $\chi_h^e$. Open red circles show $\chi_h^e$. The inset shows $\langle O^z(r)\rangle$ (red cirles) versus $r$ at $h[111]=0.25$, $\phi=0$.
  {\bf (b)}   $\langle {\cal O}^z(r=100)\rangle$ (blue triangles) versus $\phi/\pi$ at fixed $h[111]=0.15$.
  Results in {\bf (a), (b)} are from high precision iDMRG calculations with a unit cell of 60.
  The dotted vertical lines denote the transitions at critical fields determined from divergences in $\chi_h^e$ and $\chi_h^{\lambda_1}$.
  }
  \label{fig:AFSOP}
\end{figure}

Results for ${\cal O}^z$ are shown in Fig.~\ref{fig:AFSOP}. In the presence of a non-zero magnetic field or a non-zero $\Gamma$ the string
order correlation function ${\cal O}^z(r)$ is not long-range. Instead, as shown in the inset in Fig.~\ref{fig:AFSOP}(a) for $h[111]=0.25$, it
decays exponentially to zero. However, the length scale describing this exponential decay is extremely large, often exceeding hundreds or
for small enough $\Gamma$, $h[111]$, thousands of lattice spacings. The extent of the \AK\ phase can therefore be determined by
determining ${\cal O}^z(r=100)$ or ${\cal O}^z(r=900)$ which remain non-zero through out the \AK\ phase. This is illustrated in Fig.~\ref{fig:AFSOP}(a)
where both ${\cal O}^z(r=100)$ (solid blue triangles) and ${\cal O}^z(r=900)$ (solid green circles) are plotted versus $h[111]$ along side $\chi_h^e$  (open red circles)
at $\phi=0$. Clearly, ${\cal O}^z(r=100)$ drops abruptly towards zero at the phase transition between \AK\ and the SC phase. Fig.~\ref{fig:AFSOP}(b) shows
${\cal O}^z(r=100)$ versus $\phi$ at fixed $h[111]=0.15$ again abrupt changes in ${\cal O}^z(r=100)$ are observed at the boundary of the \AK\ phase.

\subsubsection{Mapping to KQ model}
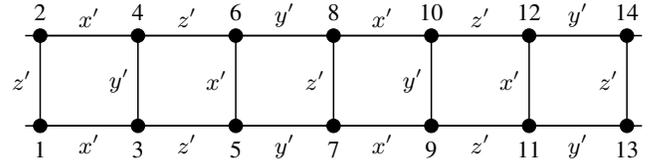
\begin{figure}[h!]
  \centering
  \begin{tikzpicture}[scale=1]
    \kgusix{6}
   \end{tikzpicture}
  \caption{The KG ladder after the $U_6$ local transformation}
  \label{fig:kgusix}
\end{figure}

It is of considerable interest to explore non-local unitary operators that will lead to string order correlation functions showing
long-range order also for $\Gamma\neq 0$. We begin by considering how the ladder is transformed under the $U_6$ transformation
previously described. 
If the spins on one leg of the ladder are numbered $i=1\ldots N/2$ we assign them a second label $k=(i-1)\mathrm{mod}6+1$
on the second leg we assign the label $k=(i+2)\mathrm{mod}6+1$.
With this labelling we introduce the following notation for the transformed bonds
\begin{eqnarray}
  x'\ :\ -KS^x_iS^x_j-\Gamma(S^y_iS^y_j+S^z_iS^z_j)\nonumber\\
  y'\ :\ -KS^y_iS^y_j-\Gamma(S^x_iS^x_j+S^z_iS^z_j)\nonumber\\
  z'\ :\ -KS^z_iS^z_j-\Gamma(S^x_iS^x_j+S^y_iS^y_j)
\end{eqnarray}
Here $j$ is a nearest neighbor site.
With this notation we can represent the transformed ladder using the following picture, Fig.~\ref{fig:kgusix}

\begin{figure}[h!]
  \centering
  \begin{tikzpicture}[scale=1]
    \wkgusixw{6}
   \end{tikzpicture}
  \caption{$H_{KQ}$: The KG ladder after the $U_6$ transformation followed by the $W$ transformation}
  \label{fig:wkgusixw}
\end{figure}
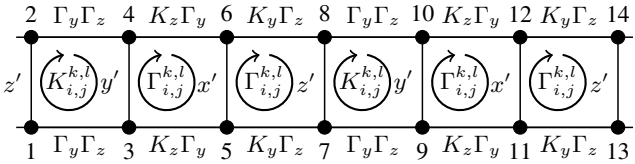

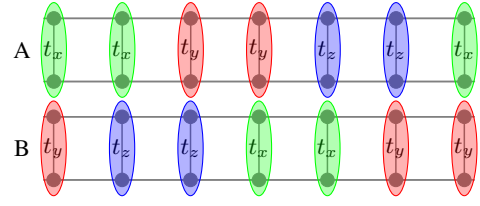
\begin{figure}[h!]
  \centering
  \begin{tikzpicture}[scale=0.7]
    \rungtripletA{6}
   \end{tikzpicture}
  \begin{tikzpicture}[scale=0.7]
    \rungtripletB{6}
   \end{tikzpicture}
  \caption{The rung-triplet states A and B for the KQ model}
  \label{fig:rts}
\end{figure}

We then consider the zero field case with OBC and change notation slightly compared to Ref.~\cite{Catuneanu2018ladder}
by using an equivalent {\it non-local} unitary operator:
\begin{equation}
 W = \prod_{\substack{j+1< k \\ j\ \mathrm{odd},\  k\ \mathrm{odd} \\ j=1,\ldots N-3 \\ k=3,\ldots N-1}} w(j,k).
\end{equation}
With the individual $w(j,k)$ given as follows:
\begin{equation}
  w(j,k)  = e^{i\pi(S^y_j+S^y_{j+1})\cdot(S^z_k+S^z_{k+1})},
\end{equation}
and $W^\dagger=W$.
With this definition of $W$ we see that on the vertical bonds of the ladder 
$W$ leaves all interactions unchanged
\begin{equation}
  WS_1^\alpha S_{2}^\alpha W = S_1^\alpha S_{2}^\alpha , \ \ WS_3^\alpha S_{4}^\alpha W = S_3^\alpha S_{4}^\alpha, \ \ldots
\end{equation}
However, on horizontal bonds we find
\begin{equation}
  WS_2^yS_4^yW = -S_1^yS_4^y, \ \ WS_3^yS_5^yW = -S_4^yS_5^y, \ \ldots
\end{equation}
Note that $W$ effectively {\it moves} the bond and changes the sign of the interaction.
Likewise, we get for the horizontal $zz$ bonds
\begin{equation}
VS_2^zS_4^zV=-S_2^zS_3^z, \ \ VS_3^zS_5^zV = -S_3^zS_6^z, \ \ldots
\end{equation}
However, the horizontal $xx$ bonds give rise to non-trivial 4-spin interactions. Specifically:
\begin{equation}
VS_2^xS_4^xV=-\sigma_1^yS_2^z\sigma_3^zS_4^y, \ \ VS_3^xS_5^xV=-S_3^z\sigma_4^yS_5^y\sigma_6^z,\ldots
\end{equation}
thereby coupling the four spins around a plaquette.
We then introduce additional notation for transformed bonds
\begin{eqnarray}
& & K_z\Gamma_y\ :\ KS^z_iS^z_j+\Gamma S^y_iS^y_j\nonumber\\
& & K_y\Gamma_z\ :\ KS^y_iS^y_j+\Gamma S^z_iS^z_j\nonumber\\
& & \Gamma_y\Gamma_z\ :\ \Gamma S^y_iS^y_j+\Gamma S^z_iS^z_j\nonumber\\
& &  \Gamma^{k,l}_{i,j}\ :\ 4\Gamma(S^z_kS^z_lS^y_iS^y_j+S^y_kS^y_lS^z_iS^z_j)\nonumber\\
& & K^{k,l}_{i,j}\ :\ 4K(S^z_kS^z_lS^y_iS^y_j+S^y_kS^y_lS^z_iS^z_j)
\end{eqnarray}
With this notation in hand
we can now apply the $W$ transformation to the $U_6$ transformed ladder shown in Fig.~\ref{fig:kgusix}.
The resulting Hamiltonian can be drawn in the manner shown in Fig.~\ref{fig:wkgusixw}.
It is quite remarkable that the $W$ transformation has generated 
four-spin exchange terms. 
We call this $H_{KQ}$ Hamiltonian the KQ ladder since 
the model is a close cousin of
JQ models~\cite{Sandvik2007} extensively studied as models of deconfined criticality~\cite{Senthil2004}.
Unexpectedly, the ground-state for the KQ model in the \AK\ phase has a significant overlap with a rung-triplet state.
If we define:
\begin{eqnarray}
  t_x = (|\uparrow\uparrow\rangle-|\downarrow\downarrow\rangle)/\sqrt{2}\nonumber\\
  t_y = (|\uparrow\uparrow\rangle+|\downarrow\downarrow\rangle)/\sqrt{2}\nonumber\\
  t_z = (|\uparrow\downarrow\rangle+|\downarrow\uparrow\rangle)/\sqrt{2}
\end{eqnarray}
Then we can pictorially draw the rung-triplet state as shown in Fig.~\ref{fig:rts}.
With OBC there is another energetically equivalent rung-triplet state obtained by translation as shown in Fig.~\ref{fig:rts}. The two states approximates the doublet ground-state of the KQ model.
Surprisingly, the unit cell for these triplet states are 12 sites and not as one might have expected from the structure of $H_{KG}$, 6 sites. 
Ordering of this type has previously been studied using SOP's inspired by the studies of $s=1$ spin chains. If $\tau_i^\alpha=S^\alpha_{i,1}+S^\alpha_{i+1,2}$ are the
sum of two diagonally situated spins, one defines~\cite{Kim2000,Fath2001}:
\begin{equation}
  {\cal O}_\mathrm{even}^\alpha(r) = -\langle \tau_i^\alpha\exp(i\pi\sum_{l=i+1}^{i+r-1}\tau_l^\alpha)\tau_{i+r}^\alpha\rangle.
  \label{eq:oeven}
\end{equation}
This order parameter has been used to distinguish topologically distinct phases in Heisenberg ladders and is non-zero in the rung-singlet phase
where the topological number is even.

\begin{figure}[t]
  \centering
        \includegraphics[width=0.5\textwidth]{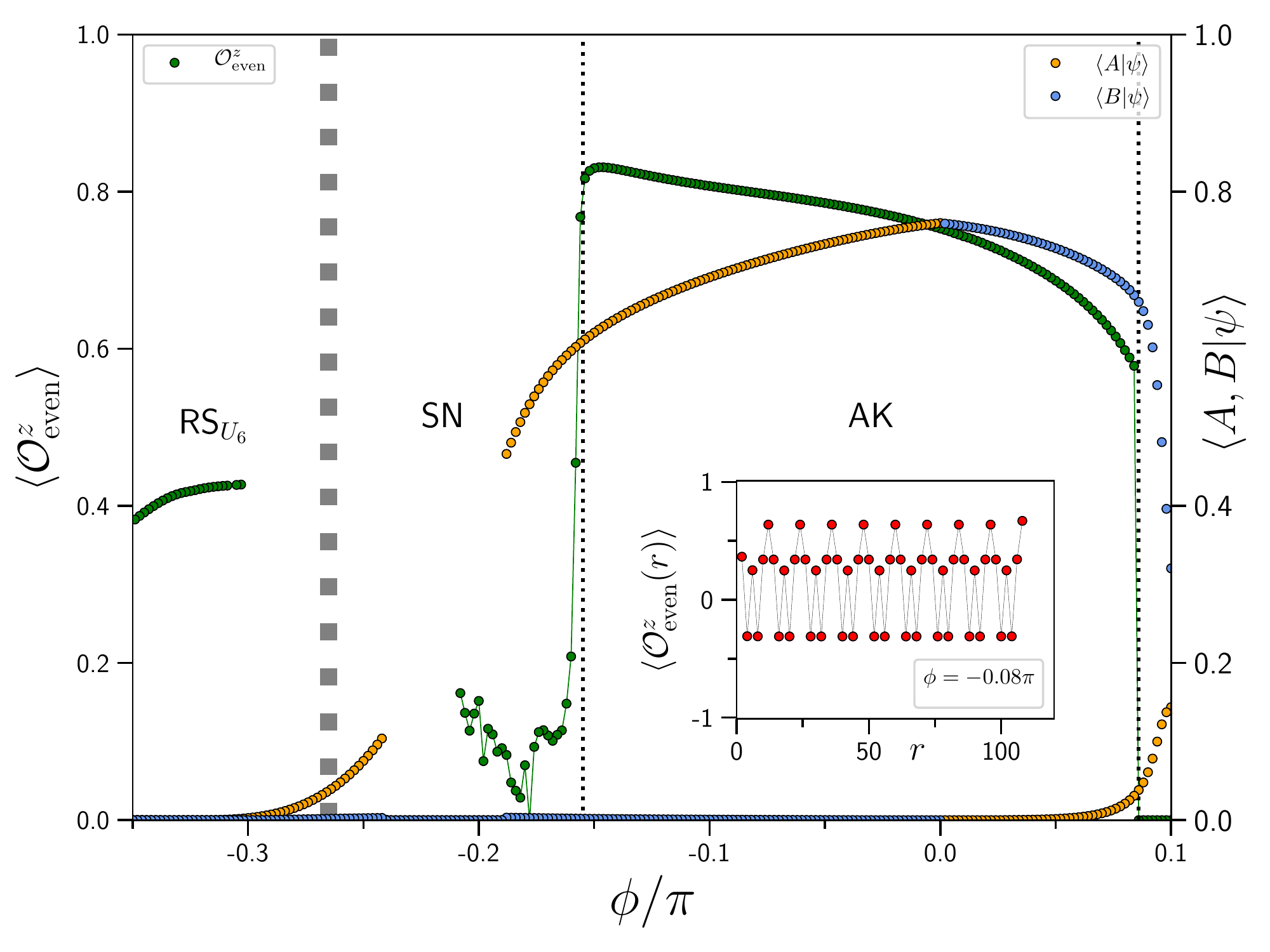}
  \caption{String order parameter $\langle {\cal O}_\mathrm{even}^z\rangle$ versus field (green circles) in the AFM Kitaev region for $H_{KQ}$ with $N=120$.
  Results are shown along side the overlaps $\langle A|\psi\rangle$ and $\langle B|\psi\rangle$ from calculations with $N=12$.
  Results are from high precision DMRG calculations with OBC.
  The dotted vertical lines denote the transitions at critical fields determined from divergences in $\chi_h^e$ and $\chi_h^{\lambda_1}$.
  }
  \label{fig:EvenS1Z}
\end{figure}

In Fig.~\ref{fig:EvenS1Z} we show results for ${\cal O}_\mathrm{even}^z$ estimated from ${\cal O}_\mathrm{even}^z(r)$ in the large $r$ limit on systems with $N=120$ and OBC.
This SOP drops abruptly to zero at the limits of the \AK\ phase and, as can be seen from the inset, there is no exponential decay observed in ${\cal O}_\mathrm{even}^z(r)$ which
instead quickly attains a constant value. As expected, ${\cal O}_\mathrm{even}^z$ is also non-zero in the RS$_{U_6}$ phase, but clearly zero in the \SN\ phase. Also shown
in Fig.~\ref{fig:EvenS1Z} are the overlaps with the rung-triplet states $\langle A|\psi\rangle$ and $\langle B|\psi\rangle$ as obtained from small $N=12$ systems with OBC. 
There are significant finite-size effects at the boundaries of the \AK\ phase for the overlaps, although they drop to zero outside the \AK\ phase the critical points are
not as well defined as for ${\cal O}_\mathrm{even}^z$. The rung-triplet states shown in Fig.~\ref{fig:EvenS1Z} are approximate and for $N>12$ different linear combinations
enter such that with the simple definitions of the rung-triplet states above $\langle A|\psi\rangle$ and $\langle B|\psi\rangle$ tend to zero as $N\to\infty$. However, ${\cal O}_\mathrm{even}^z$
is clearly non-zero in the \AK\ phase and in the KQ model this phase is therefore topologically equivalent to the Haldane like phases observed in Heisenberg ladders with
even topological number.  Since the KQ model is related to the KG model through unitary transformations the same must be true for the KG ladder throughout the \AK\ phase.
Calculations in the \FK\ phase show that ${\cal O}_\mathrm{even}^z$ is also clearly non-zero in zero field throughout that phase but drops to zero in the \AG\ phase.

\section{Comparison: the honeycomb KG model}\label{sec:compare}
It is important to note that the pure AFM Kitaev under the magnetic field studied by DMRG and 24-site ED exhibits an intermediate gapless phase before it
polarizes. A U(1) spin liquid was suggested for this field-induced gapless phase~\cite{hickey2019visons}.
However, in the two-leg ladder,  we found {\it five} distinct phases including \AK\, SC,  $\gamma$, $\eta$, and PS 
at the pure AFM Kitaev point (white thin line at $\phi=0$ in Fig. \ref{fig:FineAFK}), as the magnetic field increases. 
There are several factors including the obvious geometry difference that may result in the different results between the 24-site honeycomb cluster and current results.
To understand possible origins of the difference, we investigate the following systems at AFM Kitaev point, $\phi=0$.

\begin{figure}
  \centering
        \includegraphics[width=0.5\textwidth]{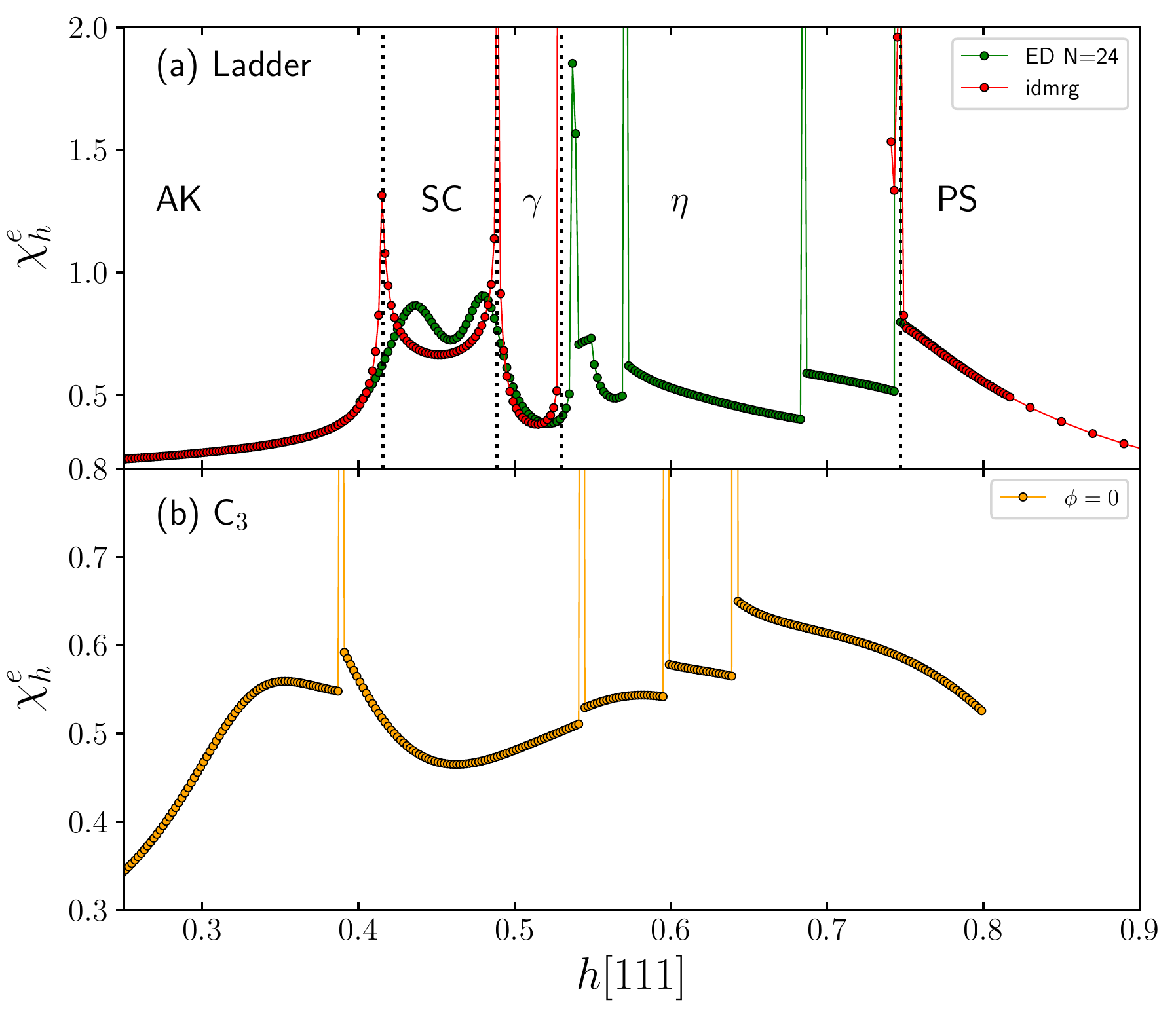}
  \caption{Comparison between the two leg ladder and the 24-site C$_3$ symmetric honeycomb geometry at $\phi=0$.
  {\bf (a)} The ladder: $\chi_h^e$ versus $h[111]$ as obtained from ED with N=24 (green circles)  and from high precision idmrg with a unit cell of 60 (red circles).
  {\bf (b)} The C$_3$ symmetric honeycomb geometry: $\chi_h^e$ versus $h[111]$ as obtained from ED with N=24 (orange circles). The critical fields are at $0.388$, $0.543$, $0.598$ and $0.641$.
  }
  \label{fig:C3Compare}
\end{figure}

 First we study the 24-site ladder using ED and compare the result with iDMRG with a unit cell of 60, to understand the finite size effects.
  The results of $\chi_h^e$ are shown in Fig. 19 (a) where the red and green dots are obtained by iDMRG and ED, respectively.
Note that, due to problems with convergence there are no iDMRG results in the $\eta$ phase.
  The sharp transitions seen in the iDMRG between \AK\ and SC at $h[111]=0.416$, and between SC and $\gamma$ at $h[111]=0.489$ are replaced by broad bumps in ED,
  while the transition between $\gamma$ and $\eta$ at $h[111]=0.530$ in the iDMRG is also sharp in the ED results. The transition to the PS phase occurs around the same field strength for both ED and iDMRG, $h[111]=0.747$.
  There are a couple of sharp features within $\eta$ phase only found in ED, which we assign to finite size effects.
  Other than these additional peaks in $\eta$ phase, the results are remarkably similar. 
 
 We also investigate the 24-site C$_3$ symmetric honeycomb cluster using the ED. $\chi_h^e$ is shown in Fig. 19 (b), where
 $h$ sweeps from 0 to 0.8 by $\delta h = 0.001$, much smaller steps than previous studies.\cite{hickey2019visons}
 There are four transitions found at $h[111]=0.388$, $0.543$, $0.598$ and $0.641$,  which may suggest three intermediate phases. For comparison Ref.~\cite{hickey2019visons} only find 2 transitions
 performing ED on the same 24-site C$_3$ cluster, presumably due to a larger $\delta h$.
 However, based on the finite size effects found in
 24-site ladder ED, we suspect that there are significant finite size effects in this system that makes it hard to determine whether these phases correspond to
 the SC, $\gamma$ and $\eta$ phases, or only one (SC) or two phases survive in the thermodynamic limit. 
 Indeed when the field is tilted away from the c-axis, there are only two transitions found in the 24-site C$_3$ cluster\cite{hickey2019visons}, while
the three intermediate phases -- SC, $\gamma$ and $\eta$ -- found in the iDMRG ladder persist even in the titlted field 
(see Appendix~\ref{app:tiltH}).
 Given that 
 $\gamma$ and $\eta$ are incommensurate magnetic phases, and thus sensitive to the shape of cluster, we speculate that they likely
 turn into another type of incommensurate phase confined in the C$_3$ symmetric cluster. The chirality of the SC phase may survive
 in the honeycomb cluster defined at a triangle made of nearest neighbors or next nearest neighbors of honeycomb lattice. This is an excellent topic for
 future study, as it requires a bigger size system to check the chiral correlation $C(r)$. 
 
 \begin{figure}
  \centering
  \includegraphics[width=0.45\textwidth]{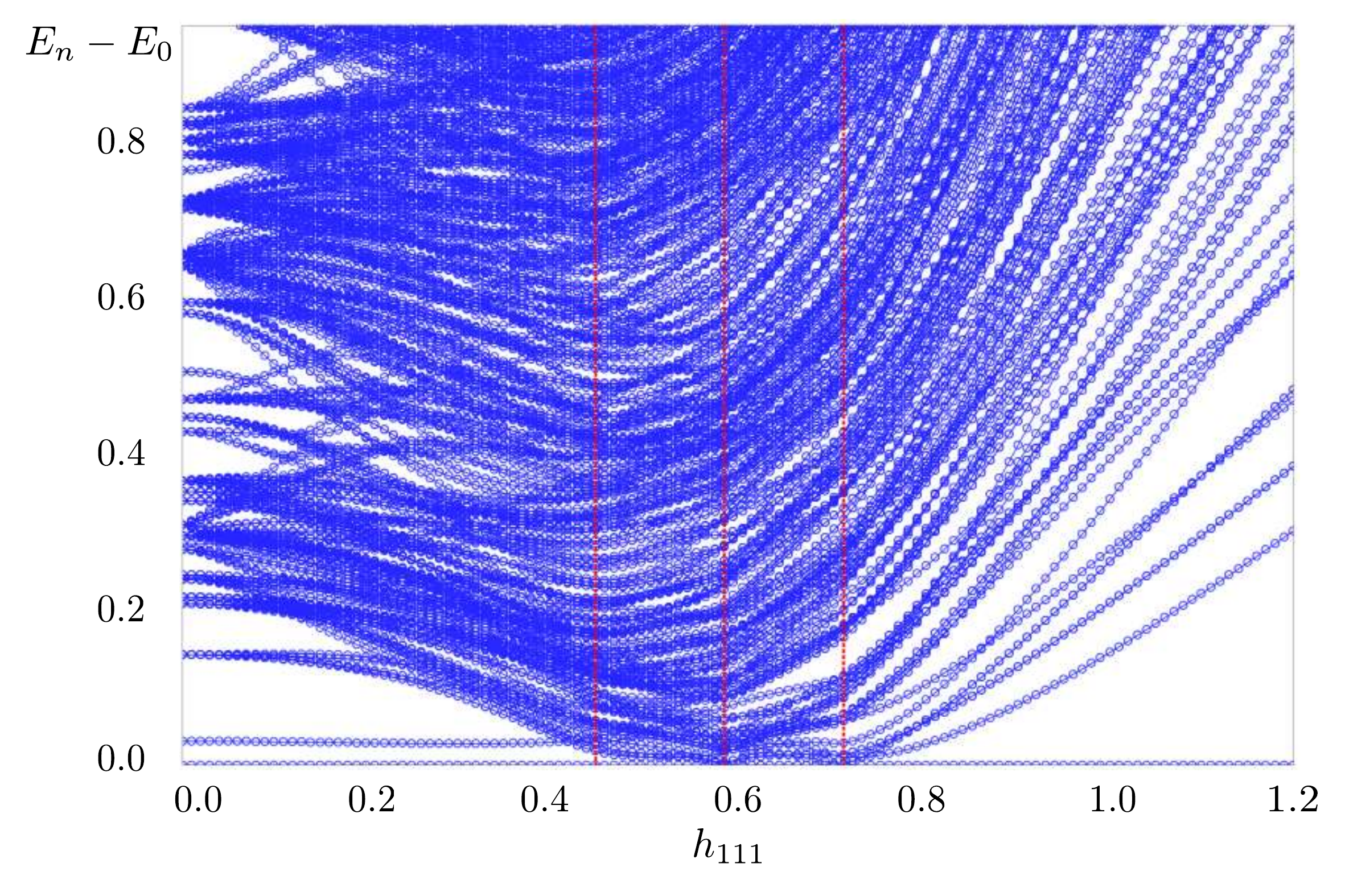} 
  \caption{The energy spectrum of the pure AFM Kitaev limit under the [111] field using 2$\times$ 6 ladder.}
  \label{fig:spec}
\end{figure}

  \begin{figure}
  \centering
  \includegraphics[width=0.5\textwidth]{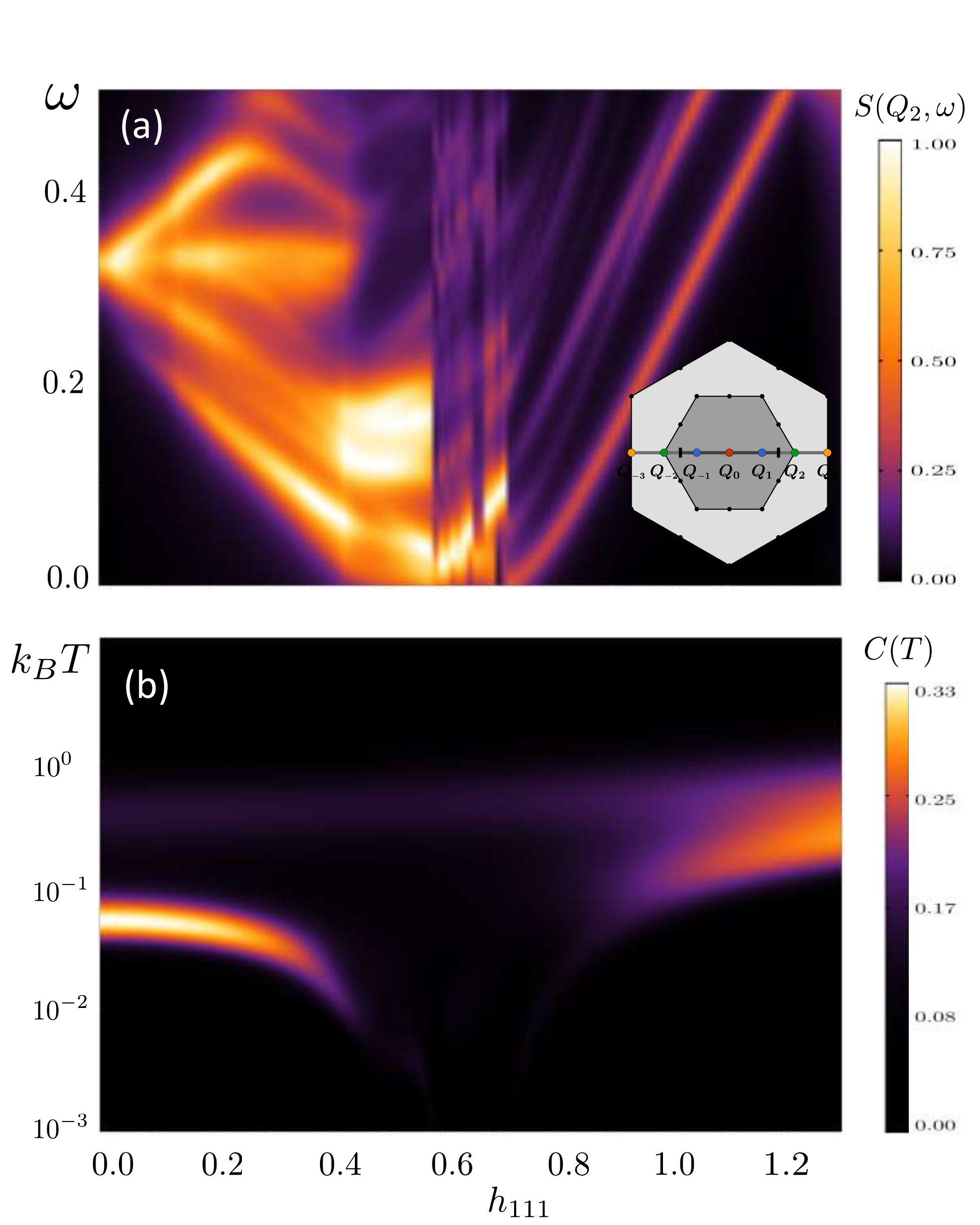}
  \caption{ (a) Dynamical spin structure factor $S(Q_2,\omega)$ at the wave-vector of Q$_2$ defined in the BZ figure shown in the inset and (b) specific heat  $C(T)$ of the pure AFM Kitaev limit under the $[111]$ field using 2$\times$ 6 ladder. }
  \label{fig:DSSF}
\end{figure}

 To shed further light on the occurrence of gapless excitations in 24-site C$_3$ cluster in ED, 
 we choose an even smaller cluster of 2 $\times$ 6 ladder, and compute various quantities under the [111] field.
The energy spectrum is shown in Fig. \ref{fig:spec} which indicates three phase transitions (red dashed lines obtained from $\chi_h^e$
 with four phases; the low-field Kitaev phase  changes to a field-induced intermediate phases, which then transitions 
 to another intermediate phase, before it becomes the polarized state. 
The collapse of the excitations is remarkably similar to 24-site C$_3$ cluster, suggesting that the intermediate gapless excitation feature is
insensitive to the cluster size and shape, even though the critical fields where such phases arise change depending on the shape.
The qualitative behaviour is similar to the 24-site ladder and C$_3$ ED presented above. However, it is not clear if there is one or two
intermediate phases, as the incommensurability of $\gamma$ and $\eta$ phases would suffer from the change of cluster size and shape, as discussed above.
The dynamical structure factor at a particular momentum $Q_2$ (shown in the inset) and the specific heat in the field
are shown in Fig. \ref{fig:DSSF} (a) and (b), respectively. The first intermediate phase is likely disordered, while the second intermediate phase
 likely exhibits incommensurate ordering. They do not exhibit well-defined excitation spectra in the specific heat similar to what is observed in the C$_3$ 24-site ED.\cite{hickey2019visons}
While it is a 12-sites ladder geometry, the qualitative results are incredibly similar to the DMRG phase diagram at the $\phi = 0$ AFM Kitaev point under the field
and  the 24-site C$_3$ symmetric honeycomb geometry results.\cite{hickey2019visons}
 It shows a dense energy spectrum in the intermediate states of both disordered (SC) and field-induced incommensurate ($\eta$ 
 or $\gamma$) phase.

 The above analysis with small ladder and C$_3$ cluster suggests that the AFM Kitaev 2D honeycomb model under the field may also 
 display a richer phase diagram than what has been reported,   and a high resolution numerical
 calculation is required to refine the phase diagram.  Since the $\gamma$ or $\eta$ phases also show a dense energy spectrum, it is important to differentiate
 the U(1) spin liquid from the incommensurate phases in the honeycomb AFM Kitaev limit. Comparing the critical field above which gapless spin liquid occurs in 24-site ED\cite{hickey2019visons}, it is likely that
 the disordered SC phase with enhanced entanglement and edge excitations extends to the gapless spin liquid in the honeycomb lattice, and
 the incommensurate phase is mixed with a polarized state, which was missed in 24-site ED of the honeycomb lattice. 
 We conclude that the ladder model at the AFM Kitaev limit captures both disordered and incommensurate magnetically ordered 
 phases under the magnetic field, and offers future directions in searching for a spin liquid in the honeycomb KG model.

\section{summary and discussion}
The KG model consists of two bond-dependent interactions, namely the Kitaev and Gamma interactions.
The Kitaev interaction on the honeycomb lattice exhibits a spin liquid with fractionalized excitations. In particular, under a time-reversal symmetry breaking term,
the excitations obey non-Abelian statistics. The Gamma interaction is another highly frustrated interaction leading to a macroscopic degeneracy in the classical limit, 
and quantum fluctuations do not lift the degeneracy found in the AFM classical Gamma model.\cite{perkins2017classical}

Since these two frustrating interactions are dominant interactions in realistic descriptions of emerging Kitaev candidates such as RuCl$_3$,
the minimal KG model was initially proposed to understand RuCl$_3$.\cite{catuneanu2018path,Gordon2019} 
The magnetic field has been a crucial parameter, as the system may undergo a transition into a field-induced
disordered phase before the trivial PS appears.
%
Aside  from its relevance to Kitaev materials, the minimal KG model may offer a playground to discover exotic spin liquids due to the combined frustration of the K and Gamma term,
and thus has been extensively studied for the last few years.
Given the huge phase space of AFM and FM Kitaev, AFM and FM Gamma, and the field,
most studies are limited to a narrow phase space focusing on the FM Kitaev and AFM Kitaev regions.
Many numerical methods have been used to identify phases of the extended Kitaev model under the field and
intriguing results were reported near AFM Kitaev and FM Kitaev regions including  
the field-induced gapless U(1) spin liquid near the AFM Kitaev region.\cite{hickey2019visons} 
However, it is not clear if the gapless excitations are due to more conventional physics such as incommensurate ordering. 

Here we investigate the entire phase space of the KG {\it ladder} model under the magnetic field.
While the geometry is limited to the ladder, it has the great advantage of allowing for high numerical precision such as accessing
iDMRG with a high precision mode with a unit cell of 60 and a maximal bond dimension of 1000.
Numerical calculations are therefore very well controlled.
We found an extremely rich phase diagram of the KG model under the field.
Among fifteen distinct phases identified, nine phases appear near the AFM Kitaev region alone.
In the zero field, there is a quadrupole ordered phase named SN, two magnetically ordered phases (FM$_{U_6}$ and RS$_{U_6}$) straightforward to understand 
from the mapping of 6-site transformation, and the disordered \AK\ phase. 
It is interesting that the SN phase found in the KG chain\cite{Wang2020b} survives in the ladder.
Other than the \AK\ phase (which becomes the Kitaev spin liquid in the 2D limit), the zero-field phases are ordered and the entanglement entropy is rather low. Under the field, highly entangled phases emerge. 
Apart from several incommensurate magnetic ordered phases, two highly entangled phases denoted by 
SC referring staggered chirality and UC uniform chirality
are induced by the field. These phases exhibit distinct chirality orderings and high entanglement entropy with gapless edge excitations 
when the boundary is open.

The ladder results presented here offer several important insights in possible spin liquids and not-yet-identified phases in the 2D honeycomb lattice. 
We would like to recall that the pure Kitaev model in the ladder corresponding to \AK\ and \FK\ in this study is gapped, 
where the ladder can be viewed as a coupled chains.\cite{Feng2007topological} As the number of chains grows, the ground state changes between gapped and gapless depending on
the even and odd numbers of the chains, and eventually maps to the 2D Kitaev spin liquid in a true 2D limit.
The \AK\ and \FK\ are gapped due to the geometry of the ladder, but its nature, magnetically disordered with high entanglement, is 
captured in the ladder model. 
Applying similar logic, we suggest that the disordered SC phase is related to the spin liquid in the 2D limit. 
The SC phase has a staggered chirality but different patterns from the \AK\ phase in field, which differentiate the two phases.
While it is gapped in the ladder, it may become gapless as the number of chains grows. 

Interestingly the previous DMRG studies on the pure AFM Kitaev point ($\phi=0$) under the $[111]$-field with three to five number of legs
reported different central charge in the intermediate phase. 
For three-leg chains, the central charge $c=1$\cite{lu2018spinon} was reported, while
for four- and five-leg chains, $c=0$\cite{lu2018spinon} and $c \sim 4$\cite{gohlke2018dynamical} respectively were found.
Based on the central charge arguments, these studies indicate that there are gapless excitations associated
 with a spinon Fermi surface in the intermediate field region. It was suggested that the spinon Fermi surface pockets are
around $K/K^\prime$- and $\Gamma$-point of the first Brillouin zone\cite{lu2018spinon},
while the other DMRG study proposed the pockets around $M$- and $\Gamma$-points.\cite{patel2019spinon}.
The existence of a spinon Fermi surface in momentum space is yet to be determined. 

While the most previous studies reported the field-induced intermediate phase as a single phase at the pure 
AFM Kitaev $\phi=0$ point in the $[111]$-field
\cite{fu2018robust,hickey2019visons,lu2018spinon,Ronquillo2019signatures,patel2019spinon,nasu2018successive}, 
a separation of the intermediate region into three phases was noted in \cite{gohlke2018dynamical},
and the middle phase, which corresponds to the $\gamma$ phase in the ladder,  
grows in extent with larger bond dimensions where five-legs were used.
We find three different intermediate phases, SC, $\gamma$ and $\eta$ in the ladder at $\phi=0$.
While the $\gamma$ phase occupies a tiny phase space in the ladder,  it is possible that this
gapless incommensurate $\gamma$ extends its phase space, as the number of chain grows. 
This implies that the chiral spin liquid candidate SC may require a finite FM $\Gamma$ interaction, as it generates more frustration. 
The SC phase with long-range chirality and enhanced entanglement appearing at the intermediate field region of the ladder
likely evolves to a field-induced spin liquid with a finite {\it staggered chirality}. 

The UC phase is another candidate of spin liquid.
It has uniform chirality pattern with high entanglement with a finite net flux,
and it appears at very low field between SN and 6-site transformed FM phase.
This phase space has not been well explored in honeycomb clusters, and we suggest further studies in this region to look for
a possible spin liquid.
The nature of the SC and UC phases and statistics of excitations in these phases are excellent topics for future study.

Possible incommensurate orderings in the 2D limit also deserves some discussion. 
Incommensurate orderings are generally difficult to pin down, as they depend on the size of cluster, 
and the ordering wave-vector itself changes even inside a phase due to the nature of incommensuration. 
In the ladder, we found several incommensurate orderings. Some has high entanglement indicating a quantum order coexisting with an incommensurate ordering.
The possibility of incommensurate ordering has been excluded in the C$_3$ symmetric cluster, mainly because of technical difficulty set by a limited size.
Our ladder study suggests several incommensurate orderings may be present in the 2D honeycomb lattice, 
and future studies on such a possibility on larger honeycomb clusters are desirable.

\begin{acknowledgments}
We have special thanks to M. Gohlke for pointing out a magnetic order in the SN phase, and 
also thank W. Yang, A. Nocera, Y. B. Kim, and I. Affleck for useful discussions.
This research was supported by NSERC and CIFAR. Computations were performed in
part on the  GPC  and  Niagara  supercomputers  at  the  SciNet  HPC
Consortium. SciNet is funded by: the Canada Foundation for Innovation under the
auspices of Compute Canada; the Government of Ontario;  Ontario  Research  Fund
- Research  Excellence; and the University of Toronto.  Computations were also
performed in part by support provided by SHARCNET (www.sharcnet.ca) and
Compute/Calcul Canada (www.computecanada.ca). Part of the numerical
calculations were performed using the ITensor library (http://itensor.org).
\end{acknowledgments}

\appendix
\begin{figure*}[t!]
  \includegraphics[width=\linewidth,clip]{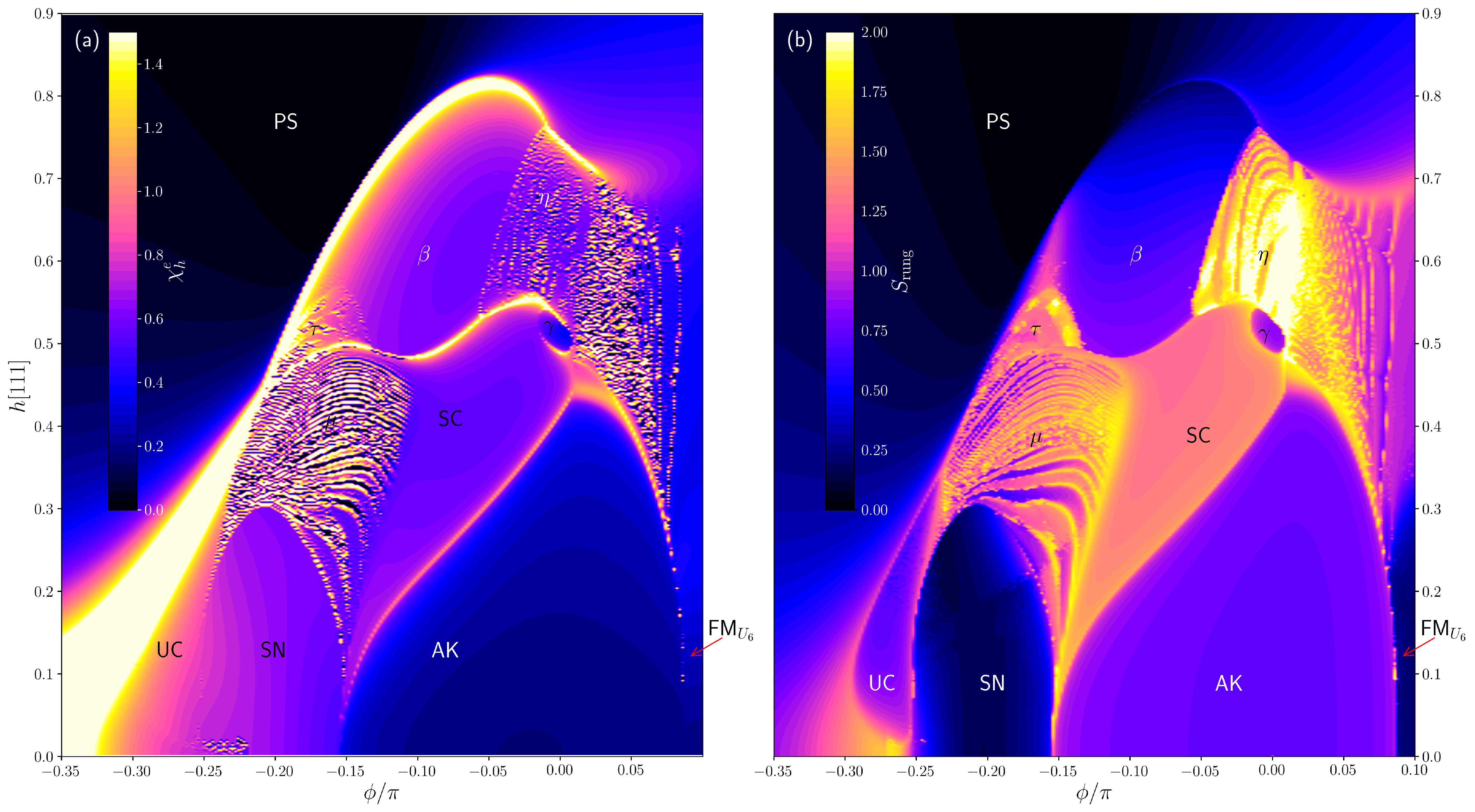} 
  \caption{
    ({\bf a}) Phase diagram for the AFM Kitaev region of the two-leg ladder KG model under the magnetic field along [111]-axis as obtained
    from $\chi^e_h$. Note the absence of a clear definition of the UC phase.
    ({\bf b}) Phase diagram for the AFM Kitaev region of the two-leg ladder KG model under the magnetic field along [111]-axis as obtained
    from $S_\mathrm{rung}$, the bipartite entanglement entropy at $N/2-1$. This partition cuts the middle rung of the ladder. Note the clear definition of the UC phase.
    All results are from high throughput iDMRG calculations with a unit cell of 24 sites, with $\Delta\phi = 0.002\pi$, $\Delta h[111] = 0.002$
  }
  \label{fig:chih}
\end{figure*}
\begin{figure}[b!]
 \includegraphics[width=\linewidth,clip]{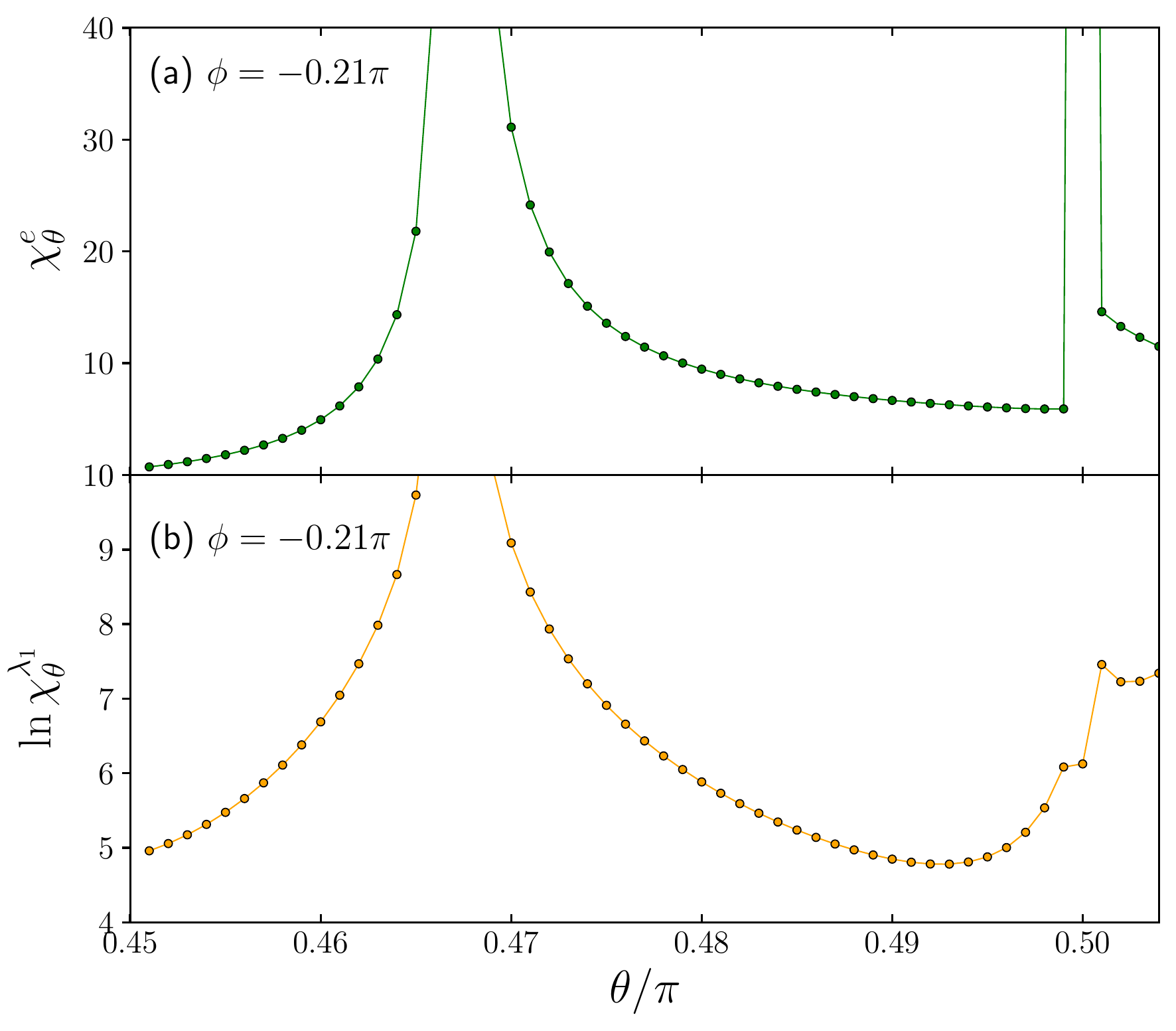} 
  \caption{
   (a) $\chi_\theta^e$ and (b) $\ln\chi_\theta^{\lambda_1}$ versus $\theta$ at a fixed $\phi = -0.21 \pi$ inside the SN phase
   where  we set $J \equiv K \cos{\theta}$ for the KG ladder.
   It shows a clear first order transition at $J=0$ indicating a first order transition point.
    Results are from high precision iDMRG with unit cell of 60.
  }
  \label{fig:JSweep}
\end{figure}

%

\section{Phase diagram of the AFM region from, $\chi^e_h$ and $S_\mathrm{rung}$}\label{app:chiesrung}
In addition to the phase diagram determined from $-\ln\lambda_1$ and $\chi^{\lambda_1}_\phi$ shown in Fig. 4 of the main text it is useful to also
consider $\chi^e_h=-\partial ^2 e_0/\partial h^2$, with $e_0$ the ground-state energy per spin. This is shown in Fig.~\ref{fig:chih}. All the phases except for the
UC phase are clearly visible in Fig.~\ref{fig:chih}(a). As described in the main text, the transition to the UC phase is subtle and easily missed in $\chi^e_h$.
However, the precursor `bump' to this phase transition (see Fig. 9(a) of the main text) is clearly 
visible in the lower left corner of Fig.~\ref{fig:chih}(a) as the large band of bright yellow, however, this feature is not
associated with any real phase transition.

Another useful depiction of the AFM Kitaev region phase diagram can be obtained from $S_\mathrm{rung}$. With $\rho_{N/2-1}$ the reduced density matrix of
the first $N/2-1$ sites of the ladder, the bipartite entanglement entropy is obtained as $S_\mathrm{rung}=-\mathrm{Tr}\rho_{N/2-1}\ln\rho_{N/2-1}$, with the partition
cutting the middle rung (and both legs) of the ladder. Our results for this quantity are shown in Fig.~\ref{fig:chih}(b). In this case the UC phase along with all the other phases are clearly defined.
Regions of increased values of $S_\mathrm{rung}$ are visible as bright yellow colored bands in the $\mu$ and $\tau$ phases. These bands likely describe lock-ins to 
particular magnetic orderings compatible with the unit cell used in the calculations. The 'heart of entanglement', with its elevated entanglement, encompassing the $\mu$, SC and $\gamma$ phases is beautifully illuminated 
in yellow and orange colors. Even higher bipartite entanglement is present in the $\eta$ phase with its bright yellow colors.

\begin{figure}[t]
  \includegraphics[width=\linewidth,clip]{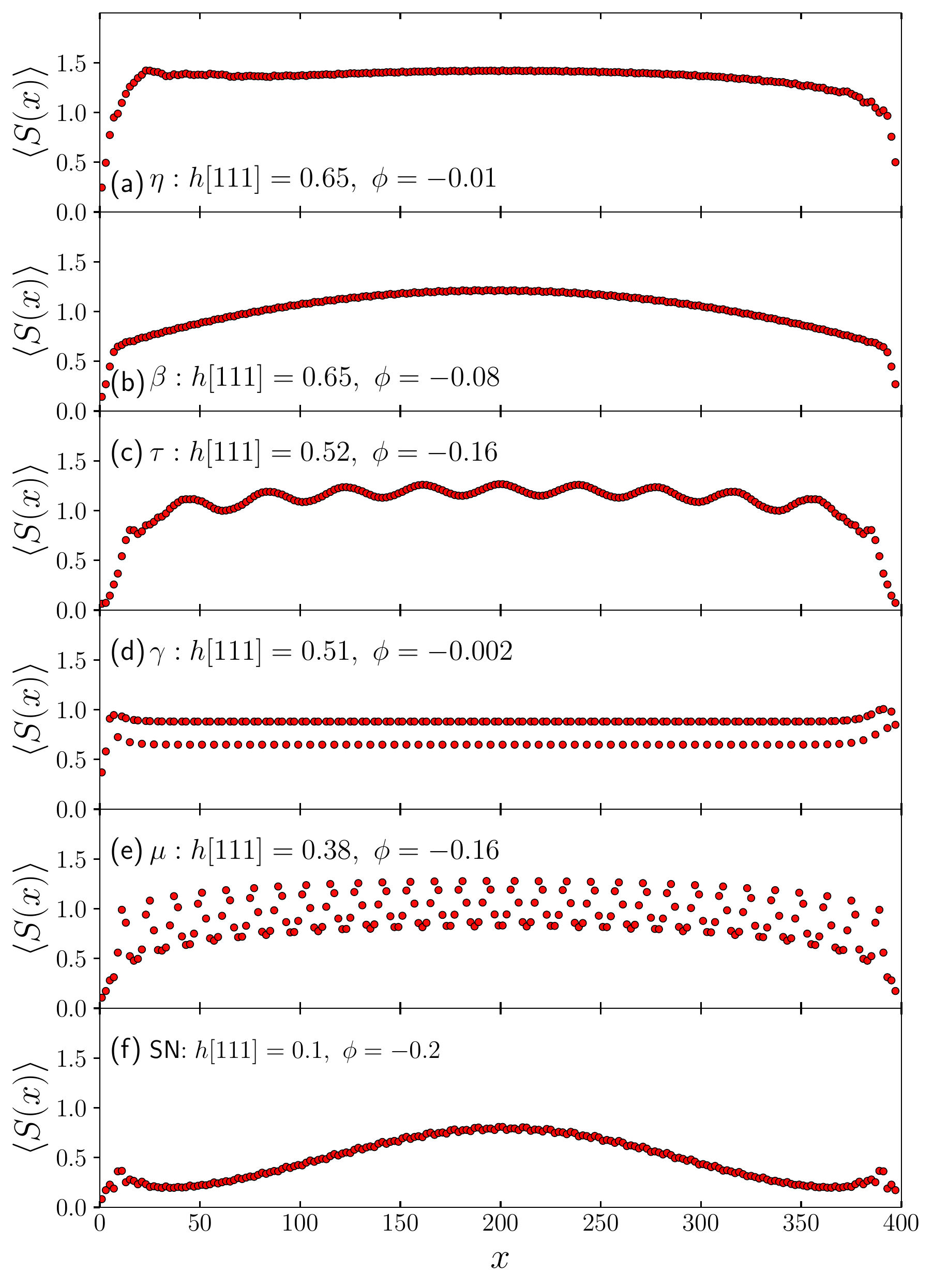} 
  \caption{
          The bi-partite rung entanglement entropy from a cut at site $x$ (odd), $S_\mathrm{rung}(x)$ versus $x$
    at six different points in the phase diagram indicated in Fig.~4 (main text) representing
  the (a) $\eta$, (b) $\beta$, (c) $\tau$, (d) $\gamma$, (e) $\mu$ and (f) \SN\ phases.
  Results are from finite-size DMRG calculations for total system size $N=400$ with open boundary conditions. 
  }
  \label{fig:AFK1EE}
\end{figure}

\begin{figure}[t]
  \includegraphics[width=\linewidth,clip]{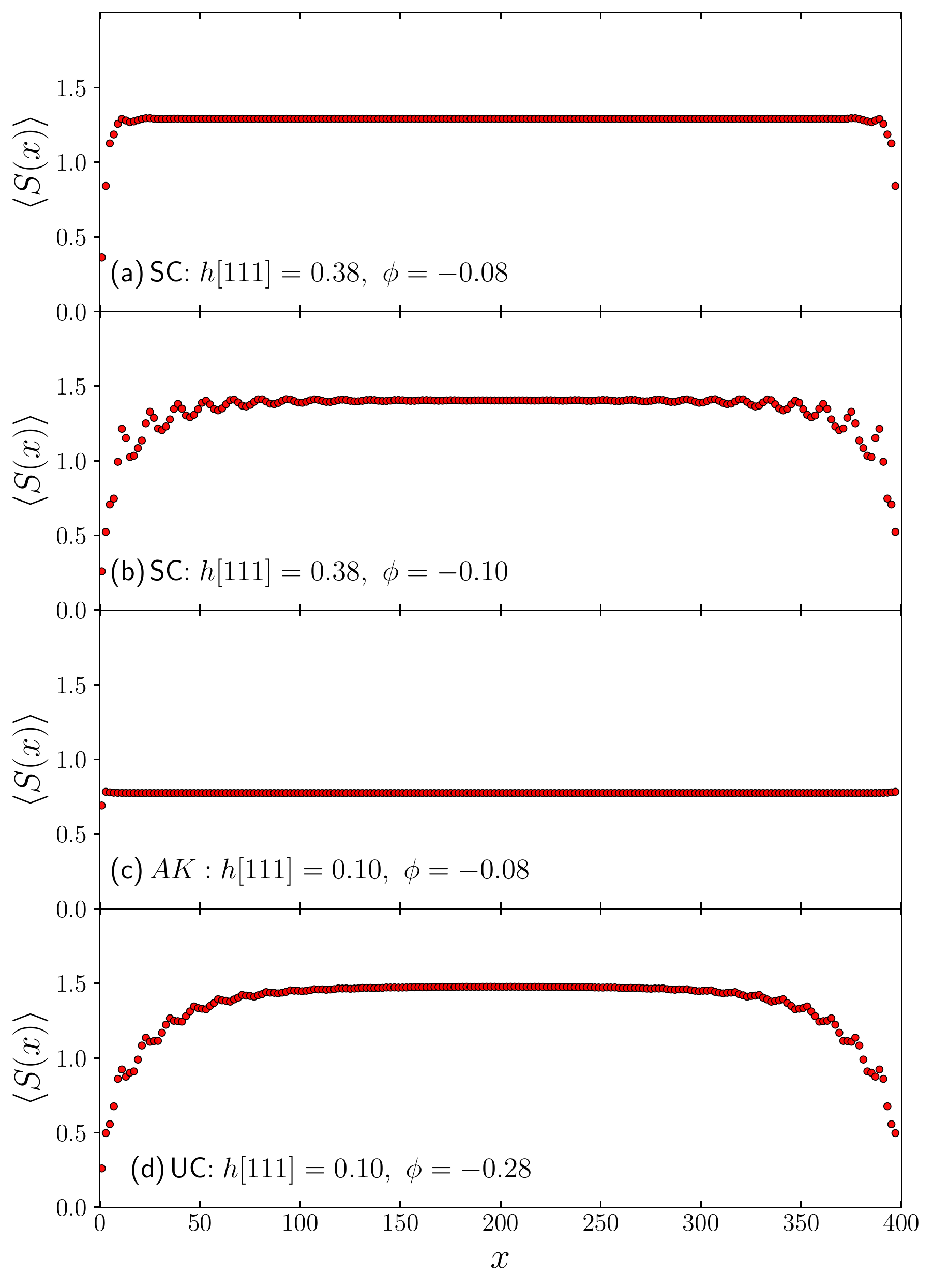} 
  \caption{
          The bi-partite rung entanglement entropy from a cut at site $x$ (odd), $S_\mathrm{rung}(x)$ versus $x$
          at four different points in the phase diagram indicated in Fig.~4 (main text) representing
  (a) the SC-phase at $h[111]=0.38$, $\phi=-0.08\pi$, (b) the SC-phase at $h[111]=0.38$, $\phi=-0.10\pi$, (c) the \AK-phase and (d) the UC phase.
  Results are from finite-size DMRG calculations for total system size $N=400$ with open boundary conditions. 
  }
  \label{fig:AFK2EE}
\end{figure}

\section{SN phase at zero field}\label{app:SN}
To understand the zero-field SN phase where the DMRG with OBC finds two different magnetic orderings
depending on the system, we introduce an additional Heisenberg coupling, $J = K \cos{\theta}$, and 
study (a) the energy susceptibility $\chi_\theta^e$ and (b) $\ln\chi_\theta^{\lambda_1}$ versus 
$\theta$ at a fixed $\phi = -0.21 \pi$.
This is to check if there is a first order transition as a function of $J$ occuring along the line of the SN.
Indeed a clear first order transition is found in the energy suscepbility at $\theta = \pi/2$, i.e. $J=0$ as shown in Fig.~\ref{fig:JSweep}. 
We conclude that the phase space denoted by the SN is a line of 
first order transitions separating two different magnetic ordering states. 
Since the nematic (spin-quadropole) order is finite in both ordered states, we keep the name the SN 
for this line of first order transitions, throughout the paper.

\section{Bipartite entanglement in the AFM Kitaev region}\label{app:sx}
If we instead of defining the reduced density matrix at $x=N/2-1$, as we did when considering $S_\mathrm{rung}$, but instead at a general $x$ always taken to be
{\it odd}, 
we obtain the bipartite rung entanglement entropy, $S_\mathrm{rung}(x)=-\mathrm{Tr}\rho_x\ln\rho_x$. Our results for $S_\mathrm{rung}(x)$ 
at various points in the AFM Kitaev region are
shown in Fig.~\ref{fig:AFK1EE} and \ref{fig:AFK2EE} corresponding to the same points as the ones shown in Fig. 5, 6 in the main text. 
It is well established that the entanglement entropy is bounded~\cite{Hastings2007,Gottesman2010} in systems with a gap. For a system with
a gap $S_\mathrm{rung}(x)$ should then attain a plateau in the middle of the chain for long enough ladders when that limit is attained.
On the other hand, for a 1D critical gapless system with open boundary conditions we expect~\cite{Cardy04}
\begin{equation}
  S(x)=\frac{c}{6}\ln\left[\frac{2N}{\pi}\sin\left(\frac{\pi x}{N}\right)\right]+\ln g+s_1/2.
  \label{eq:sx}
\end{equation}
Here $c$ is the central charge, $g$ the universal ground-state degeneracy~\cite{Affleck1991} and $S_1$ a non-universal constant.
As is well known, the entanglement therefore grows logarithmically and is not bounded.
The fact that a plateau is observed in the middle of the ladder for $S_\mathrm{rung}$ is therefore consistent with a gap.

From the results shown in Fig.~\ref{fig:AFK1EE} and \ref{fig:AFK2EE} the behavior of $S_\mathrm{rung}(x)$ in the SC, AK, UC and $\gamma$ phases
are therefore indicative of a gapped phase. The results for the remaining phases are more consistent with a gapless phase, although without giving
a clear fit to the form Eq.~(\ref{eq:sx}) and it is not possible to extract the central charge.

\begin{figure}[t]
  \includegraphics[width=\linewidth,clip]{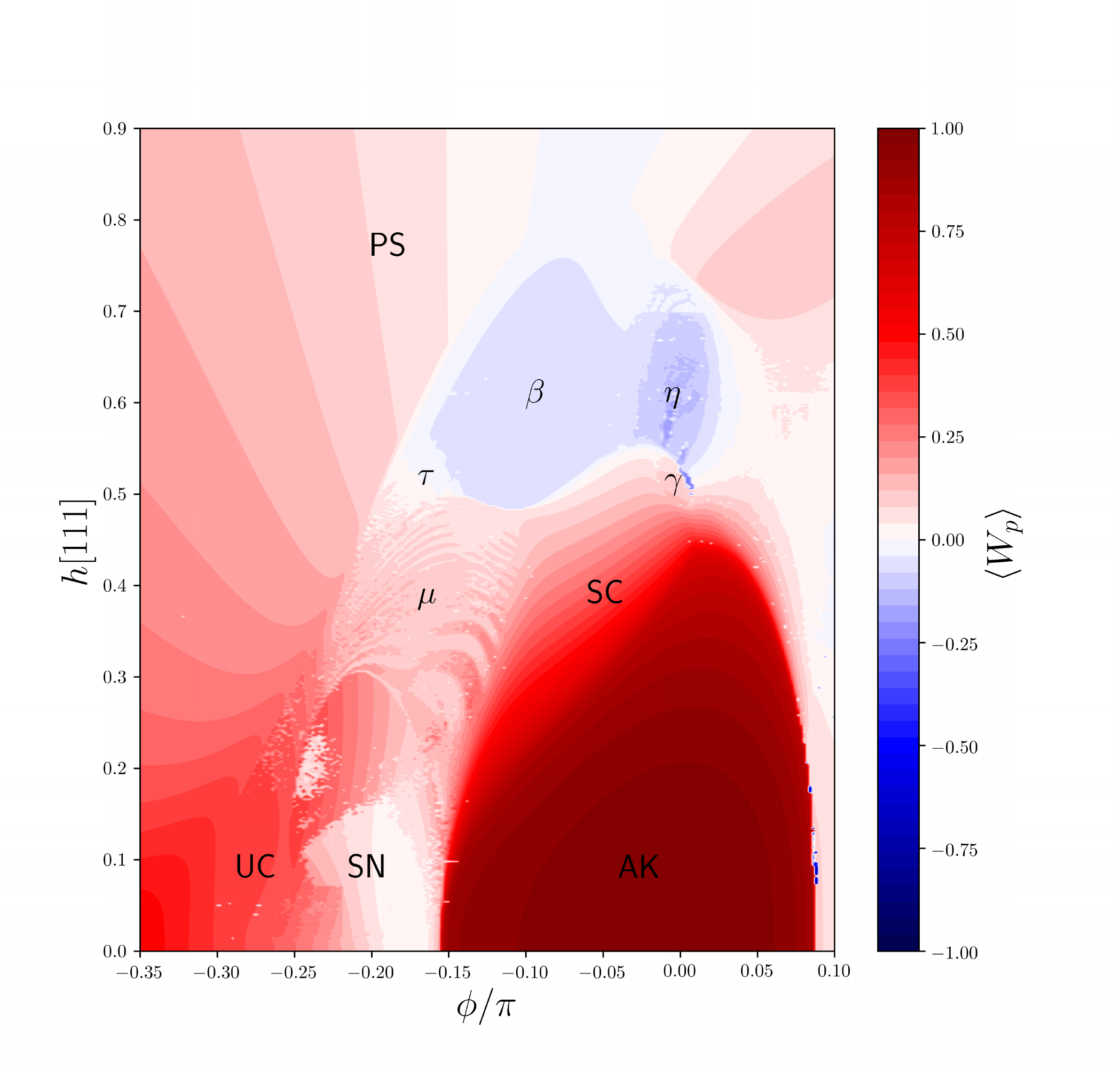} 
  \caption{
    The honeycomb plaquette operator, $W_p$, through out the AFM Kitaev region for the two-leg ladder KG model under the magnetic field along [111]-axis.
    Note the negative values in the $\beta$ and $\eta$ phases.
    All results are from high throughput iDMRG calculations with a unit cell of 24 sites, with $\Delta\phi = 0.002\pi$, $\Delta h[111] = 0.002$.
    The white speckles are missing data points.
  }
  \label{fig:Wp}
\end{figure}

\begin{figure}[t]
  \includegraphics[width=\linewidth,clip]{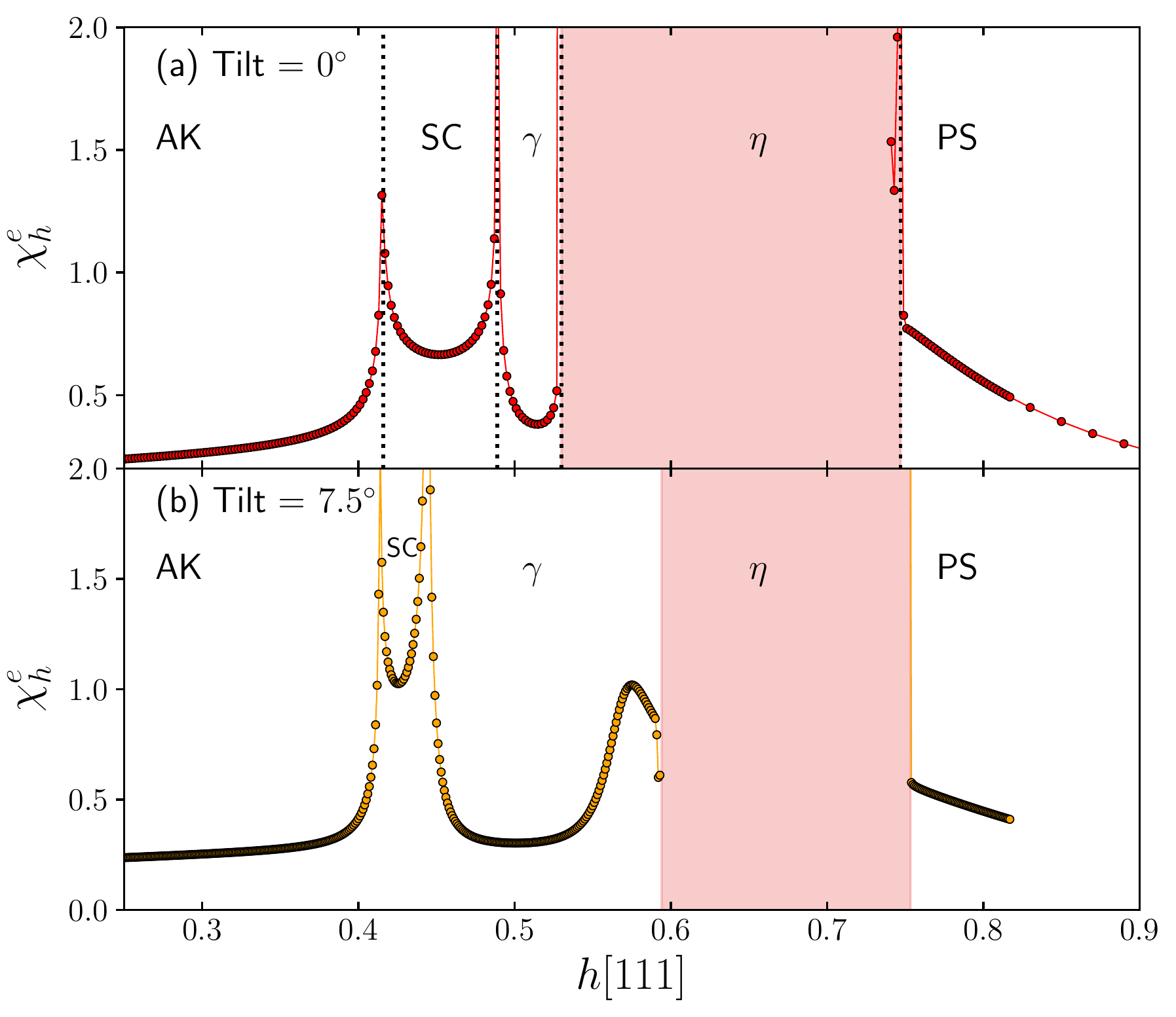} 
  \caption{
    $\chi_h^e$ versus field for the KG ladder for (a) $h[111]$ field and (b) a field $\cos(\theta)h[111]+\sin(\theta)h[11-2]$ with $\theta=7.5^\circ$.
    The light red shading indicates regions where the iDMRG does not converge well.
    Results are from high precision iDMRG with unit cell of 60.
  }
  \label{fig:Tilt}
\end{figure}

\section{The plaquette operator, $W_p$, in the AFM Kitaev region}\label{app:wp}
The plaquette operator, $W_p$, introduced by Kitaev~\cite{kitaev2006}, is a crucial tool for characterizing the Kitaev spin liquid. On the honeycomb lattice $W_p$ is defined 
as a product of the six spins surrounding a hexagon in the lattice. In the setting of the KG ladder derived
from a strip of the two-dimensional honeycomb lattice $W_p$ then corresponds to the 6 spins around 2 plaquettes of the ladder. 
Following the numbering of the sites on the ladder shown in Fig. 16 in the main text we then have:
\begin{equation}
  W_p=\sigma_1^y\sigma_2^z\sigma_3^x\sigma_4^x\sigma_5^z\sigma_6^y.
\end{equation}
Our results for $\langle W_p\rangle$ are shown in Fig.~\ref{fig:Wp}.
At both the AFK and FM Kitaev points, $\phi=0$ and $\phi=\pi$ we have a flux-free state with $\langle W_p\rangle =1$ on all plaquettes. Away from 
these points $W_p$ can be different from 1. However, as can be seen in Fig.~\ref{fig:Wp}, $\langle W_p\rangle$ remains high throughout the AK phase
only decreasing when the phase is exited. However, the different phases in the AFK region are only partly visible in Fig.~\ref{fig:Wp}. Interestingly,
$\langle W_p\rangle$ is negative in the $\beta$ and $\eta$ phases shown as blue in Fig.~\ref{fig:Wp}.

\section{AF Kitaev point in Tilted Field}\label{app:tiltH}
It is an interesting question if the phases found around the AF Kitaev point persist when the field direction is changed. In order to investigate this
we have calculated $\chi_h^e$ in the presence of a field tilted slightly away from the [111] direction towards the [11-2] direction of the following 
form  $\cos(\theta)h[111]+\sin(\theta)h[11-2]$ with $\theta=7.5^\circ$. Our results, obtained from iDMRG are shown in Fig.~\ref{fig:Tilt}. In panel
Fig.~\ref{fig:Tilt}(a) we show for comparison our previous results for $\theta=0$ and in Fig.~\ref{fig:Tilt}(b) for a tilted field with $\theta=7.5^\circ$.
The AK, SC and $\gamma$ phases are still clearly present as is the transition to the polarized phase PS. However, the transition to the $\eta$ phase is
in this case less well defined. In both panels the light red coloring indicates regions where the iDMRG does not converge well.
Remarkably the critical fields for the AK-SC and $\eta$-PS transition appear almost unchanged by the tilt of the field.
\bibliography{references}

\end{document}